\documentclass[final,5p,times,twocolumn,centertitle]{elsarticle}

\usepackage{amssymb}
\usepackage{amsmath}

\usepackage{booktabs}
\usepackage{multirow}


\usepackage[labelfont=bf]{caption}

\biboptions{sort&compress}

\journal{Journal of Magnetism and Magnetic Materials}

\begin{document}

\begin{frontmatter}



\title{Anisotropic Magnetic and Transport Properties of RAlGe (R = Y, Gd--Tm, Lu)}


\author[inst1,inst2]{Tyler J. Slade} 

\author[inst1]{Lin-Lin Wang} 

\author[inst1,inst2]{Sergey L. Bud'ko} 

\author[inst1,inst2]{Paul C. Canfield} 

\affiliation[inst1]{organization={Ames National Laboratory},
            addressline={}, 
            city={Ames},
            postcode={50011}, 
            state={Iowa},
            country={USA}}

\affiliation[inst2]{organization={Department of Physics and Astronomy, Iowa State University},
            addressline={}, 
            city={Ames},
            postcode={50011}, 
            state={Iowa},
            country={USA}}

\begin{abstract}
We grew single crystals of the \textit{R}AlGe family for \textit{R} = Y, Gd-Tm, Lu and characterized them with powder X-ray diffraction, temperature dependent specific heat measurements as well as temperature and field dependent resistance and magnetization measurements. These \textit{R}AlGe materials crystallize in an orthorhombic \textit{Cmcm} crystal structure in which the $R$ atoms occupy a position with $m$2$m$ point symmetry. We find that when \textit{R} is a moment bearing rare earth atom, the \textit{R}AlGe materials order antifferomagnetically at \textit{T}$_N$ ranging from 5 K (TmAlGe) to 39 K (TbAlGe). A second, lower temperature antiferromagnetic transition is also detected in the range 6--8 K for \textit{R} = Tb--Ho, but no lower transition is observed (for T $\geq$ 1.8 K) for \textit{R} = Er, Tm. The $R$ = Tb-Tm members show evidence for strong crystal field effects influencing the physical properties, where the crystal field favors an easy $a$-axis for $R$ = Tb--Ho and an easy $b$-axis for $R$ = Er, Tm. Of these compounds, strongly uniaxial behavior is observed for DyAlGe and TmAlGe. At \textit{T} = 2 K, rich metamagnetism is observed when the external field is applied along the \textit{a}-axis for \textit{R} = Tb--Ho, whereas a single metamagnetic transition is observed (up to 70 kOe) with the field along the \textit{b}-axis when \textit{R} = Er and Tm. All \textit{R}AlGe materials show metallic transport behavior and have moderate positive magnetoresistance up to 60--150$\%$ at 2 K superimposed on top of metamagnetic features. The field dependent magnetization measurements on non-moment bearing YAlGe show clear de Haas–van Alphen oscillations, indicating the Fermi surface includes small, high mobility pockets.
\end{abstract}




\begin{keyword}



rare earth magnetism \sep topological semimetals  \sep intermetallics

\end{keyword}

\end{frontmatter}



\section{Introduction}

The \textit{R}AlGe (\textit{R} = rare earth element) materials exhibit an attractive combination of unique magnetotransport properties and potentially flexible topological states. The light rare-earth (\textit{R} = La--Gd) members of the series crystallize in a noncentrosymmetric, tetragonal \textit{I}4$_1$\textit{md} structure, and following early predictions that suggested the broken inversion symmetry may enforce type-II Weyl nodes \cite{xu2017discovery,chang2018magnetic}, were found to exhibit a wealth of interesting transport and magnetic phenomena \cite{flandorfer1998systems,dhar1992magnetic,hodovanets2018single,puphal2019bulk,piva2023topological,hodovanets2022anomalous,suzuki2019singular,meng2019large,wang2020correlation,dhital2023multi,yang2023stripe,kikugawa2024anomalous,cho2023large,liu2021critical,puphal2020development,zhao2022field}. Highlights include the potential observation of a topological Hall effect in CeAlGe \cite{piva2023topological}, and high anomalous Hall and Nernst effects that are in agreement with values predicted from the intrinsic Berry curvature in PrAlGe and NdAlGe \cite{meng2019large,dhital2023multi,yang2023stripe,kikugawa2024anomalous,cho2023large}. CeAlGe also displays complex angular magnetoresitance \cite{hodovanets2022anomalous}, and when alloyed with Si, CeAlGe$_{1-x}$Si$_x$, the magnetoresistance undergoes enormous spikes exceeding 1000$\%$ when the applied field is aligned within 1° of the \textit{a}- or \textit{b}-axis \cite{suzuki2019singular}. This unusual behavior was suggested to arise from an interaction between magnetic domain walls and the small Fermi surface associated with the Weyl nodes.

\begin{figure}
    \centering
    \includegraphics[width=0.6
    \linewidth]{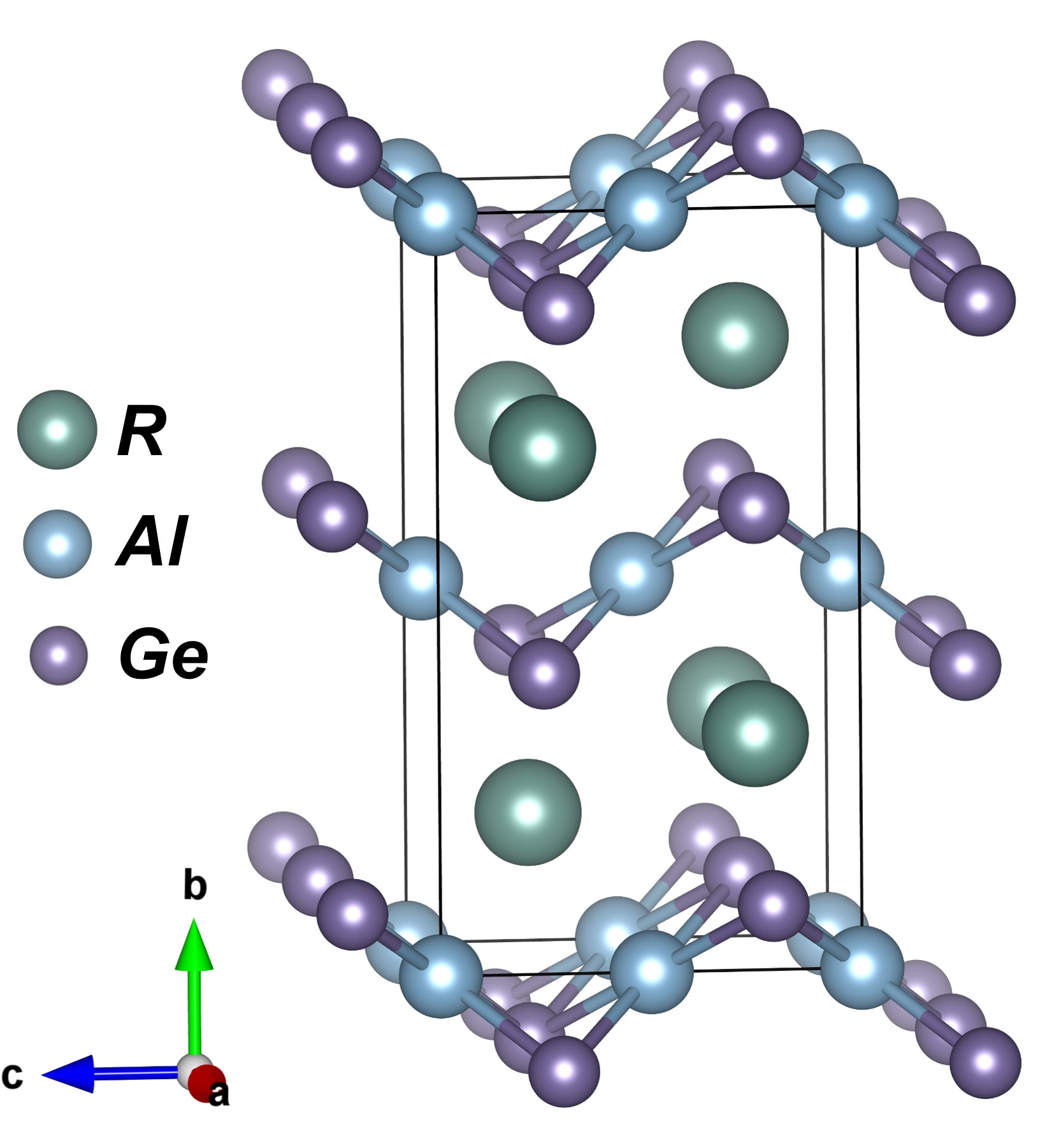}
    \caption[]{Crystal Structure of orthorhombic, \textit{Cmcm}, \textit{R}AlGe, where \textit{R} = Y, Gd--Tm, Lu. The solid black lines denote the unit cell.}
    \label{Structure}
\end{figure}

The rare earth (\textit{R} = Y, Gd--Tm, Lu) based \textit{R}AlGe compounds crystallize in a different, orthorhombic, crystal structure with space group \textit{Cmcm} \cite{zhao1990structure}, except for GdAlGe, which can adopt either a high temperature tetragonal \textit{I}4$_1$\textit{md} or low temperature \textit{Cmcm} polymorph. Fig. \ref{Structure} shows that the crystal structure of the orthorhombic \textit{Cmcm} varient consists of corrugated Al-Ge layers and alternating triangles of \textit{R} atoms that fill the space between the Al-Ge layers. The Al-Ge layers are composed of a sheet of Al atoms that span the \textit{ac} plane and which are linked above and below, in an alternating manner, by Ge atoms that form chains extending along the \textit{a}-axis. The \textit{R} atoms occupy a position with relatively low $m$2$m$ point symmetry in the space between the layers, sitting at the vertices of alternating isosceles triangles related to each other by inversion symmetry.

Compared to the light rare earth based \textit{R}AlGe compounds, the heavy \textit{R}AlGe materials remain relatively understudied. Existing data indicate antiferromagnetic ordering occurs between $\approx$ 7--40 K for \textit{R} = Gd--Er, \cite{wang2021magnetic,wang2021correlation,wang2022dynamic,gouda2022angular,kumar2026anisotropic} but to the best of our knowledge, no data concerning the physical properties of TmAlGe have been published. Very recent work on GdAlGe suggests the presence of strongly dispersive, Dirac-like bands that support a substantial Fermi velocity on the order of 10$^6$ m/s, much greater than typically observed in magnetic topological semimetals. \cite{laha2025single} 

Here, we grow single crystals of \textit{R}AlGe for \textit{R} = Y, Gd--Tm, Lu and provide characterization of their magnetic behavior, electronic transport properties, and specific heat. We find all samples (besides YAlGe and LuAlGe) order antiferromagnetically, with \textit{T}$_N$ between 5 K (for TmAlGe) up to 39 K (for TbAlGe). When \textit{R} = Gd, Tb, Dy, and Ho, additional lower temperature antiferromagnetic transitions are detected under 10 K, whereas \textit{R} = Er, Tm members do not show signatures of a second transition down to 1.8 K. For \textit{R} = Tb--Ho, field dependent magnetization measurements at 2 K show multiple metamagnetic transitions when the field is applied along the \textit{a}-axis. A single, \textit{T} = 2 K, metamagnetic transition is detected when the field is along the \textit{b}-axis for \textit{R} = Er and Tm. All samples have metallic resistivity and show clear anomalies associated with the magnetic transitions. The heavy \textit{R}AlGe compounds generally have moderate positive magnetoresistance on the order of 60--150$\%$ depending on the \textit{R} atom, with peaks or anomalies corresponding their metamagnetic transitions. Density functional theory calculations suggest the electronic structure of nonmagnetic \textit{R}AlGe may support Dirac nodal lines, and presence of quantum oscilation in the magnetization of YAlGe suggests the Fermi surface contains small, low-mass bands. This work therefore provides an overview of the basic physical properties of the \textit{R}AlGe family for heavy rare earths (\textit{R} = Gd-Lu, except Yb) and serves as a basis for more detailed investigation of each member.

\section{Experimental details}

\begin{figure*}[!t]
    \centering
    \includegraphics[width=0.8\linewidth]{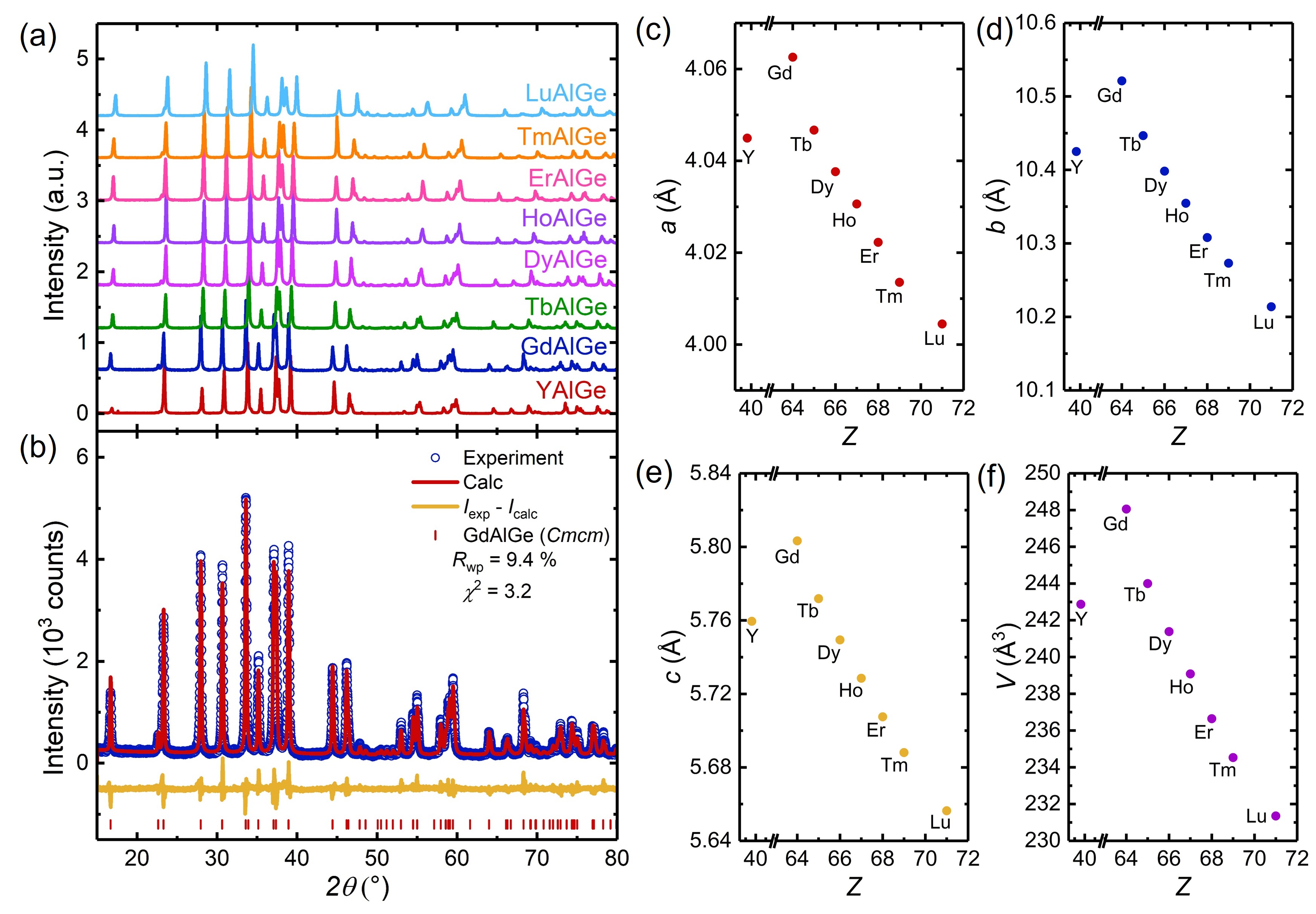}
    \caption{(a) Powder X-ray diffraction patterns for the \textit{R}AlGe samples. The patterns are normalized to the highest intensity (1 3 0) peak and offset for clarity. (b) Rietveld refinement of GdAlGe using the \textit{Cmcm} structural model. (c)-(e) Refined lattice parameters and (f) unit cell volume of \textit{R}AlGe as a function atomic number \textit{Z}. Note the x-axis break between Y and the 4\textit{f} elements (Gd-Lu). For each plot (c-f), the YAlGe data point falls between those for TbAlGe and DyAlGe.}
    \label{PXRD}
\end{figure*}

\subsection{Crystal growth} 
Single crystals of \textit{R}AlGe were grown from a high temperature solution \cite{canfield2019new} based on excess aluminum. Starting with elemental rare earths (all Ames Lab, 99.99 $\%$), Al chunks (Alfa-Aesar 99.99 $\%$), and Ge pieces (Alfa-Aesar 99.99 $\%$), the elements were weighed in a nominal molar ratio of \textit{R}$_{0.1}$(Al$_{0.86}$Ge$_{0.14}$)$_{0.9}$ and loaded into a 3 piece alumina Canfield crucible set. \cite{canfield2016use,lspceramics_CCS} The crucibles were sealed under vacuum inside of fused silica tubes, loaded into a box furnace, and heated to 1150°C. After dwelling at 1150°C for 1 h, the furnace was cooled to 700°C over 40 h, upon which the tubes were rapidly removed from the furnace, inverted, and the excess flux decanted in a centrifuge equiped with metal cups. Opening the crucibles revealed shiny, well faceted, three dimensional crystals with lengths of $\approx$ 3--6 mm. Thus far, our attempts to produce a hypothetical YbAlGe using the same or similar procedure have not been successful. 

\subsection{X-ray diffraction} To determine the phase of the crystals, samples were analyzed with powder X-ray diffraction (PXRD). Several crystals from each batch were ground to a powder, sifted through a 35 micron mesh sieve, and diffraction patterns were collected at room temperature on a Rigaku Miniflex-II instrument operating with Cu-K$\alpha$ radiation ($\lambda$ = 1.5406 Å) at 30 kV and 15 mA. To determine the lattice parameters, the powder data was refined by the Rietveld method using GSAS-II software \cite{toby2013gsas}.

To test for possible site disorder, \cite{wang2022dynamic} additional single crystal X-ray diffraction was performed on a HoAlGe crystal using a Rigaku XtaLab Synergy-S diffractometer with Ag  radiation (0.56087 \AA\ ) operating at 65 kV and 0.67 mA. The samples were held in a nylon loop with vacuum grease, and the data was collected at room temperature. The total number of runs and images was based on the strategy calculation from the program CrysAlisPro (Rigaku OD, 2023). The data integration and reduction were also performed using CrysAlisPro, and a numerical absorption correction was applied based on Gaussian integration over a face-indexed crystal. The structures were solved by intrinsic phasing using SHELXT software and were refined with SHELXL. The crystallographic data and refinement results are given in Tables \ref{table_refinement}-\ref{table_coordinates} in the appendix, and the atomic positions and thermal displacement parameters are also listed in the appendix in Table \ref{table_Uanis}.

Prior to measuring the physical properties, the crystals were oriented using a Multiwire Laue Camera operating with X-rays from a W source ($\lambda$ = 0.7107 \AA) at 10 mV and 30 mA with $\approx$ 10 cm between the sample and detector.

\subsection{Physical property measurements} 

The magnetization measurements were performed in a Quantum Design Magnetic Property Measurement System SQUID magnetometer. The samples were mounted on a kelF disk, and a blank background using the bare disk was first measured and the values subtracted. Unless noted, all temperature dependent measurements were field cooled, and the field dependent isotherms collected on increasing \textit{H}.

The temperature and magnetic field dependent electrical resistivity were measured in a Quantum Design Physical Property Measurement System. The samples were prepared by cutting the crystals in suitable bars. The contacts were 25 $\mu$m thick annealed Pt wire that were either spot welded onto or attached with silver epoxy (EPO-TEK H20E) to the \textit{R}AlGe samples in standard four point geometry. The spot welded contacts were covered with a small portion of silver epoxy to ensure mechanical stability. Typical contact resistances were on the order of 1 $\Omega$.

Specific heat was measured in a Quantum Design Physical Property Measurement System. Prior to each measurement, an addenda was measured using the blank PPMS puck with a small amount of N-grease. For the moment bearing \textit{R}AlGe samples, the magnetic specific (\textit{C}$_{mag}$) heat was estimated by subtracting the data for LuAlGe, which served as a reference for the electronic and phonon contributions. The magnetic entropy was determined by integrating \textit{C}$_{mag}$/\textit{T}. For the integration, the specific heat data below our base temperature (1.8-1.9 K) were extrapolated with a spline fit (using Origin 2026 software) to \textit{C}$_{\text{p}}$ = 0 at \textit{T} = 0 K.

\subsection{Electronic band structure calculations}

Band structure calculations in density functional theory \cite{hohenberg1964inhomogeneous,kohn1965self} (DFT) were performed with the PBE \cite{perdew1996generalized} exchange-correlation functional using a plane-wave basis set and projector augmented wave method, \cite{blochl1994projector} as implemented in the Vienna Ab-initio Simulation Package \cite{kresse1996efficiency,kresse1996efficient}(VASP). We use a kinetic energy cutoff of 240.3 eV, a $\Gamma$-centered Monkhorst-Pack \cite{monkhorst1976special} (10 $\times$ 10 $\times$ 7) k-mesh, and a Gaussian smearing of 0.05 eV. For the accurate calculation of density of states (DOS), we use a dense (20 $\times$ 20 $\times$ 14) k-mesh with the tetrahedron method. \cite{blochl1994improved} To calculate Wilson loop with Wannier charge centers, maximally localized Wannier functions (MLWF) \cite{marzari1997maximally,souza2001maximally} and the tight-binding model were constructed with Y $sd$ , Al $sp$ and Ge $sp$ orbitals to reproduce closely the band structure around the Fermi energy (EF) within $E_{\text{F}}$ $\pm$ 2 eV. 

\begin{figure*}[!t]
    \centering
    \includegraphics[width=\linewidth]{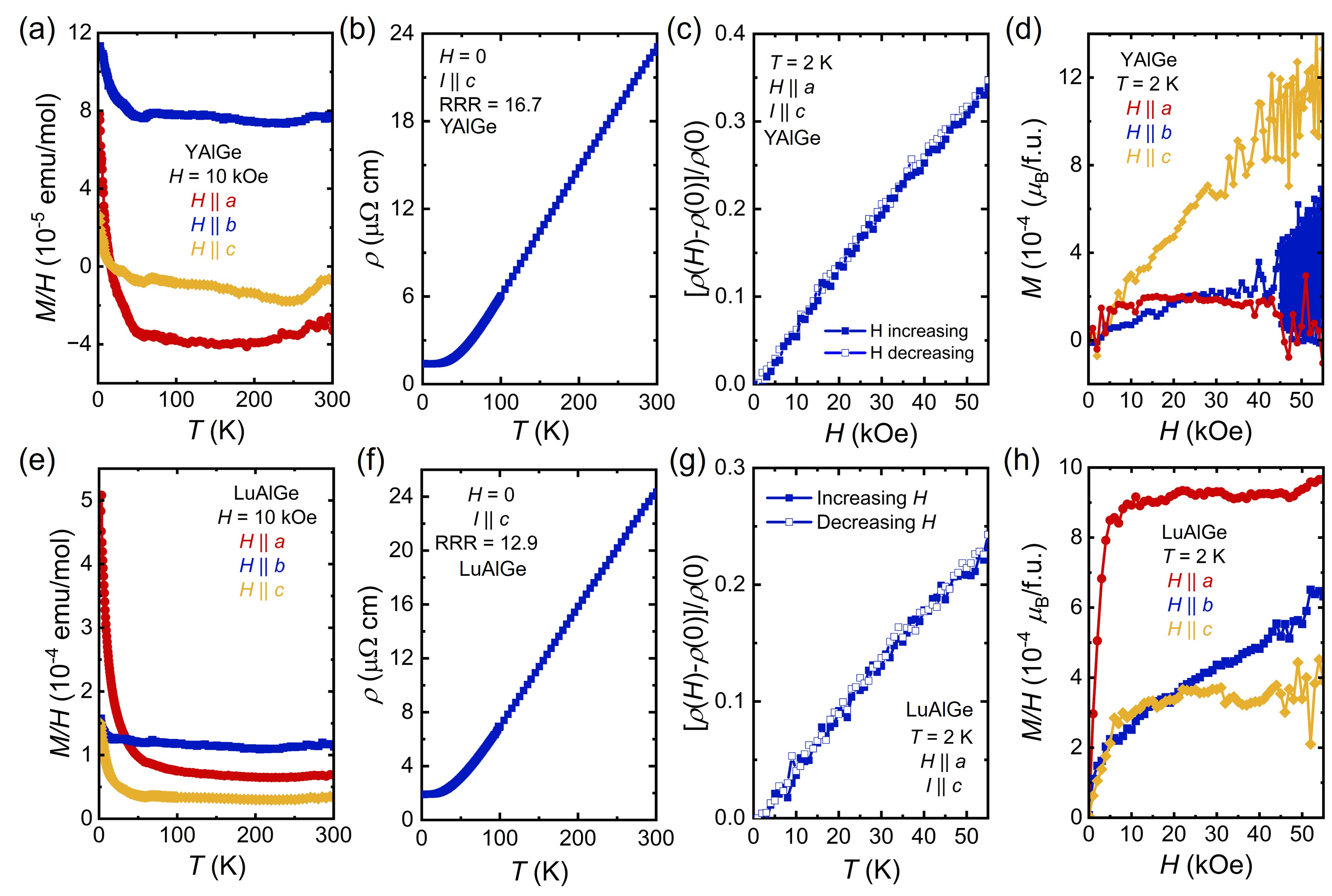}
    \caption[]{(a) Temperature dependent magnetic susceptibility (\textit{M/H}), (b) temperature dependent resistivity, (c) field-dependent resistivity (MR), and (d) Field-dependent magnetization for YAlGe. (e-h) show the same data for LuAlGe.}
    \label{YAlGe_LuAlGe}
\end{figure*}

\begin{figure}[!t]
    \centering
    \includegraphics[width=\linewidth]{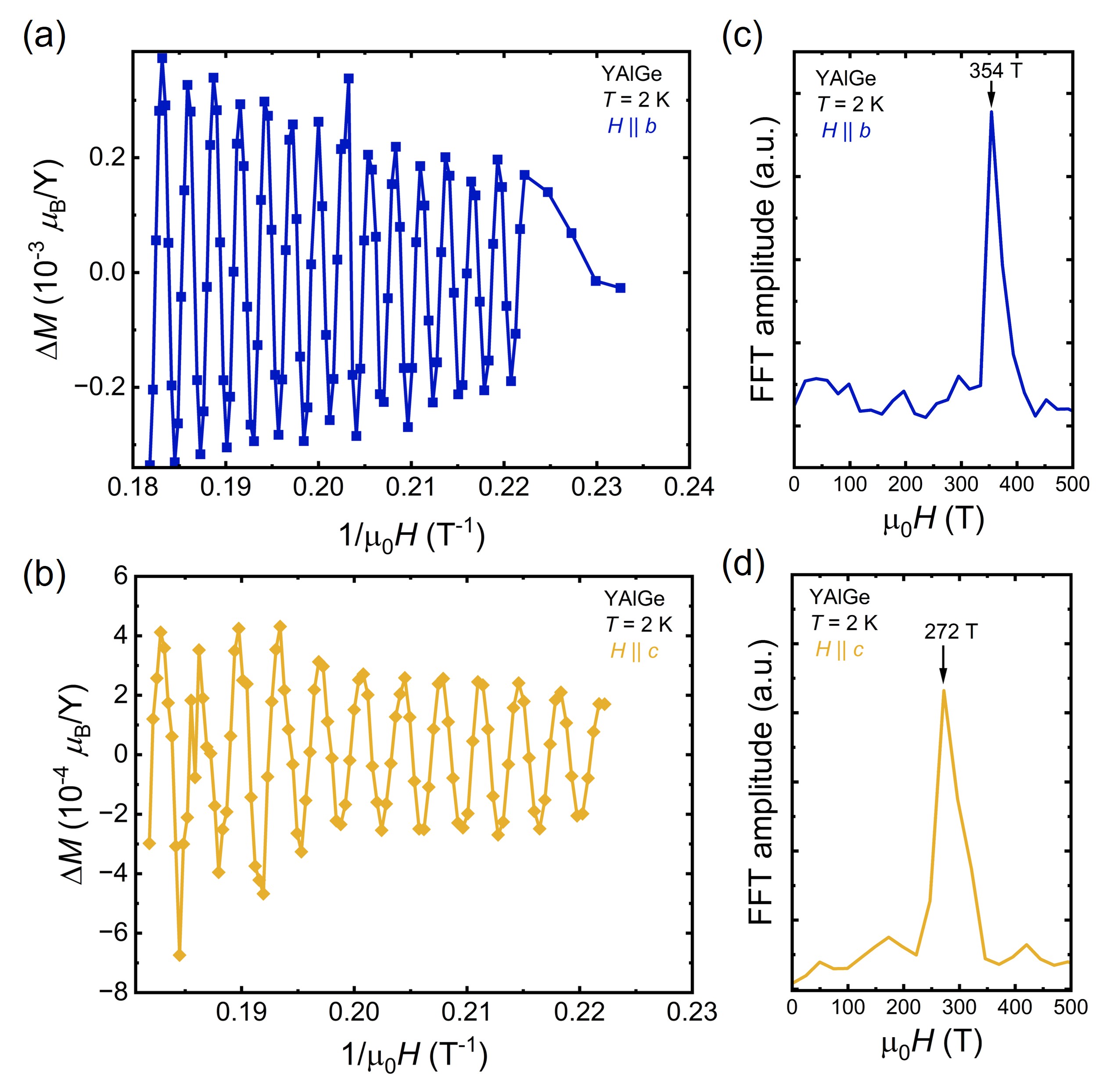}
    \caption[]{Analysis of the De Haas–van Alphen oscillations for YAlGe (a) and (b) show the oscillatory component of the magnetization plotted against inverse field for \textit{H} $\parallel$ \textit{b} and \textit{H} $\parallel$ \textit{c} orientations, respectively. (c) and (d) show the fast-Fourier transforms of the associated data in (a) and (b).}
    \label{YAlGe_QOs}
\end{figure}

\section{Results and discussion}

\subsection{Structural characterization} Fig. \ref{PXRD}a shows powder X-ray diffraction patterns obtained from the \textit{R}AlGe samples and Fig. \ref{PXRD}b shows a representative Rietveld refinement for GdAlGe (analogous refinements were performed for each sample's pattern). The observed reflections are in excellent agreement with the pattern anticipated for the \textit{Cmcm} YAlGe type structure (illustrated in Fig. \ref{Structure}), and the refinements give consistently reasonable quality fits to the experimental data, with R$_{\text{wp}}$ = 8--13 and GOF = 1.8-3.3. We note that several patterns (in particular that for YAlGe) show very weak peaks (with intensity less than 100 counts) that are most likely from Y-Al or Y-Ge binary phases that form from excess flux which was not fully decanted off the crystals. Given such week intensities attributable to secondary phases, the PXRD patterns support successful growth of (essentially) single phase $R$AlGe crystals. The refined lattice parameters for each material are given in Figs. \ref{PXRD}c-\ref{PXRD}f as a function of the atomic number \textit{Z} of the respective rare earth atoms. For \textit{R} = Gd--Tm, Lu, all three lattice parameters decrease monotonically as the rare earth element moves from Gd to Lu, in line with the expected Lanthanide contraction. The lattice parameters for YAlGe are close to, but smaller than those found for TbAlGe with the \textit{a}-lattice parameter being the closest and the \textit{c}-lattice parameter be further below, but still closer to that found for TbAlGe than for DyAlGe.

Whereas the powder diffraction data discussed above support the \textit{Cmcm} YAlGe type structure for \textit{R} = Y, Gd--Tm, Lu, a recent publication on DyAlGe and HoAlGe suggested that these two members of the heavy \textit{R}AlGe series adopt a disordered variant of this structure, with the same \textit{Cmcm} symmetry but in which the \textit{R}, Al, Ge all have mixed occupancy on each crystallographic site. \cite{wang2022dynamic} Because the disordered structural model was based on fits to powder diffraction patterns, we investigated the possibility of a disordered structure by collecting and analyzing single crystal X-ray diffraction data on our HoAlGe crystals. The results of our structure solutions and refinements are presented in Tables \ref{table_refinement}-\ref{table_Uanis} in the appendix. In short, we find no strong evidence for a disordered structure, as the fully ordered arrangement, with atomic coordinates given in Table \ref{table_coordinates} (and illustrated in Fig. \ref{Structure}) gives an excellent fit to the diffraction data, and attempts to introduce mixed occupancy onto the respective positions yielded refinements in which the occupancies of the expected atoms changed by less than 1 $\%$. Whereas we only measured single crystal diffraction on HoAlGe, the temperature dependent resistivity datasets, discussed in the following section, have residual resistance ratios, RRR = $\rho$(300 K)/$\rho$(1.8 K), in the range 10--20 for each member of the series, and we furthermore observe de Haas-van Alphen oscillations in the magnetization isotherms for YAlGe. Both the high RRR values and observed quantum oscillations point to weak carrier scattering, inconsistent with substantial site disorder, and coupled with the single-crystal X-ray diffraction data for HoAlGe, we conclude that there is no (significant) site-occupancy disorder in the heavy the \textit{R}AlGe family.

\vspace{1mm}
\subsection{Physical properties}

\begin{figure*}[!t]
    \centering
    \includegraphics[width=\linewidth]{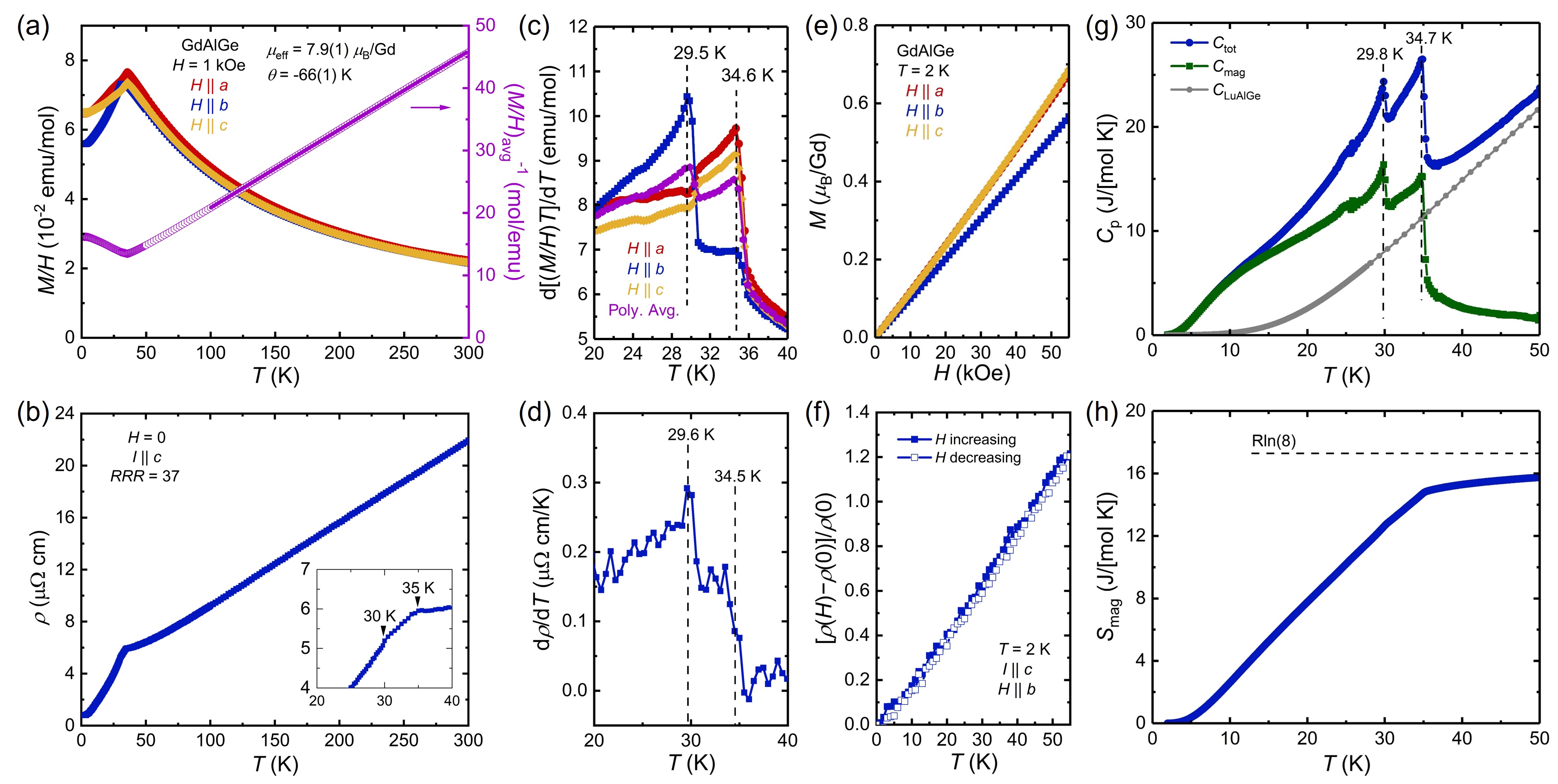}
    \caption[]{Physical properties of GdAlGe. (a) Temperature dependent magnetic susceptibility (left axis). The right axis shows the average of the individual inverse (\textit{M/H})$^{-1}$ curves for each orientation (see text). The solid line shows the Curie-Weiss fit to the ($M/H$)$_{\text{avg}}^{-1}$ data (open points). (b) Temperature dependent resistivity. The inset in (b) gives a close-up view of the data between 20--40 K. (c) Derivative of the temperature-susceptibility product (d\textit{(M/H)T}/d\textit{T}), (d) derivative of the resistivity (d$\rho$/d\textit{T}), (e) field dependent magnetization at \textit{T} = 2 K, (f) magnetoresistance at \textit{T} = 2 K, (g) total and magnetic contributions to the heat capacity. The data for LuAlGe is also shown for reference. (h) Estimated magnetic entropy.}
    \label{GdAlGe_data}
\end{figure*}

\subsubsection{YAlGe and LuAlGe}

We begin our discussion of the physical properties of the \textit{R}AlGe family with YAlGe and LuAlGe, which contain non-moment-bearing Y or Lu atoms. Fig. \ref{YAlGe_LuAlGe}a and \ref{YAlGe_LuAlGe}e show the temperature dependent magnetic susceptibility, here defined as \textit{M/H}, of YAlGe and LuAlGe measured at 10 kOe. Both compounds have small \textit{M/H}, on the order of 10$^{-5}$-10$^{-4}$ emu/mol, that is nearly temperature independent, but clearly anisotropic, above $\approx$ 50 K. For YAlGe, the magnetization is negative when the field is aligned along the \textit{a}- and \textit{b}-directions and positive along the \textit{c}-axis. This indicates the Landau and/or core diamagnetic contributions to the susceptibility are dominant in the former orientations and the Pauli paramagnetic contribution is dominant in the latter. In LuAlGe, the Pauli paramagnetism dominates the susceptibility in each sample orientation. In both YAlGe and LuAlGe, the susceptibility increases rapidly at low temperatures beginning below $\approx$50 K, which can likely be attributed to the presence of magnetic impurities in the samples.

Figs. \ref{YAlGe_LuAlGe}b and \ref{YAlGe_LuAlGe}f show the temperature dependent electrical resistivity $\rho$ for each compound. Both YAlGe and LuAlGe behave as normal metals, with resistivity that decreases approximately linearly on cooling before saturating at the lowest temperatures. The residual resistance ratios ($\rho$(300 K)/$\rho$(2 K)) are 16.7 and 12.9 for YAlGe and LuAlGe respectively, attesting to good crystal quality and low levels of disorder. The resistance data has no evidence for transitions, consistent with the magnetic data and with the expectations for non-moment bearing Y and Lu. The magnetoresistances ($\rho$(\textit{H})--$\rho${0})/$\rho$(0) are presented in Figs. \ref{YAlGe_LuAlGe}c and \ref{YAlGe_LuAlGe}g. Both compounds have similar behavior, in which $\rho$(\textit{H}) increases with a sub-linear field dependence to modest, unsaturated, values of 35 $\%$ and 24 $\%$ at 55 kOe for YAlGe and LuAlGe respectively. Magnetization isotherms measured at base temperatures of 1.8 K and 2 K are presented in Figs. \ref{YAlGe_LuAlGe}d and \ref{YAlGe_LuAlGe}h. de Haas–van Alphen oscillations are clearly present in the data for YAlGe, and their amplitude increases at higher fields, becoming most apparent above 40 kOe. No quantum oscillations are observed for LuAlGe, and instead the magnetization is weakly field dependent above 10 kOe. At lower fields, the magnetization isotherms for LuAlGe show a rapid increase, which, considering the minute (10$^{-4}$ $\mu_{\text{B}}$) values of \textit{M}, likely reflects the polarization of magnetic impurities extrinsic to the bulk sample.

The presence of quantum oscillation in the magnetization isotherms for YAlGe suggests the Fermi surface contains small/low mass and high mobility pockets. To quantify this further, Figs. \ref{YAlGe_QOs}a and \ref{YAlGe_QOs}b plot the oscillatory part of the magnetization, $\Delta$\textit{M} as a function of inverse magnetic field, where $\Delta$\textit{M} is the difference between the total measured magnetization and the non-oscillatory component, the latter of which was estimated using a third order polynomial fit to the \textit{M}(\textit{H}) isotherms. We note that to achieve appropriate data density, the data in Fig. \ref{YAlGe_QOs} was obtained from separate \textit{M}(\textit{H}) measurements than that shown in Fig. \ref{YAlGe_LuAlGe}d and were only measured between 40-55 kOe where the oscillations are strong. Figs. \ref{YAlGe_QOs}c and \ref{YAlGe_QOs}d show Fast-Fourier transforms (FFTs) of the quantum oscillations in \ref{YAlGe_QOs}a and \ref{YAlGe_QOs}b respectively. In both orientations, the FFT reveals one strong frequency, at 354 T for \textit{H} $\parallel$ \textit{b} and 272 T when \textit{H} $\parallel$ \textit{c}. From the frequencies, we calculated the associated Fermi-surface cross sections $S_{\text{F}}$ using the Lifshitz-Onsager relation:

\begin{equation}
    F = \frac{\hbar}{2\pi e}S_{\text{F}}
\end{equation}

\noindent where $F$ are the quantum oscillation frequencies corresponding to each pocket, \textit{h} is Plank's constant, and $e$ is the electron charge. The 354 T and 272 T components for \textit{H} $\parallel$ \textit{b} and \textit{H} $\parallel$ \textit{c} are both consistent with small, low mass pockets, and represent cross sections of 3.4$\times10^{-3}$ $\AA^{-2}$ and 2.6$\times10^{-3}$ $\AA^{-2}$, accounting for 1.9 $\%$ and 1.4 $\%$ of the respective $k_{\text{x}}$--$k_{\text{z}}$ and $k_{\text{x}}$--$k_{\text{y}}$ planes.

\subsubsection{GdAlGe}

The temperature dependent magnetization for GdAlGe is shown in Fig. \ref{GdAlGe_data}a. The susceptibility is effectively isotropic at high temperatures, as expected for Gd$^{3+}$ with orbital angular momentum \textit{L} = 0. \textit{M/H} decreases on cooling and reaches a peak near 35 K, indicating the onset of antiferromagnetic order. The resistivity, shown in Fig. \ref{GdAlGe_data}b, is consistent with the magnetic data, with a clear loss of spin disorder scattering observed below $\approx$ 35 K. Close inspection of the low-temperature data (insets in Figs. \ref{GdAlGe_data}a and \ref{GdAlGe_data}b) indicate there are two relatively closely spaced transitions, and based on the extrema in the respective derivatives d\textit{(M/H)T}/d\textit{T} and d$\rho$/d\textit{T} shown in Figs. \ref{GdAlGe_data}c and \ref{GdAlGe_data}d \cite{fisher1962relation,fisher1968resistive}, we assign transition temperatures of \textit{T}$_{N1}$ = 34.6 K and \textit{T}$_{N2}$ = 29.5 K, with good agreement between magnetic and transport data. The data presented here show good agreement with the previous publications on GdAlGe \cite{laha2025single}, however, the lower temperature transition was not observed in the polycrystalline samples. 

The right axis of Fig. \ref{GdAlGe_data}a shows the inverse of the average of the three \textit{M/H} curves, where (\textit{M/H})$_{\text{avg}}$ = $\frac{1}{3}$($M_{a}/H$ + $M_{b}/H$ + $M_{c}/H$). Above the Néel temperature, the inverse susceptibility is linear with temperature, consistent with thermally disordered local moments, and we therefore fit the (\textit{M/H})$_{\text{avg}}^{-1}$ to the Curie-Weiss law over 100-300 K, where

\begin{equation}
  (M/H)_{avg} = \frac{C}{T-\theta}  
\end{equation}

\noindent in which \textit{C} is the Curie constant, \textit{T} the temperature, and $\theta$ the Weiss temperature. From the fit, we estimate $\theta$ = -66(1) K, suggesting antiferromagnetic correlations, and an effective moment $\mu_{\text{eff}}$ = 7.9(1) $\mu_B$/Gd, very close to the value of 7.94 $\mu_B$ expected for Gd$^{3+}$. 

The field dependent magnetization measured at 2 K is given in Fig. \ref{GdAlGe_data}e, and the results are also consistent with an antiferromagnetic ground state. For each orientation, \textit{M} increases linearly over the measured range of 0--55 kOe. The magnetization with the field oriented along \textit{a} and \textit{c} axes are nearly identical, and the field dependence has a smaller slope with \textit{H} $\parallel$ \textit{b}, which is consistent with the smaller susceptibility, in the ordered state, for this orientation implied by Fig. \ref{GdAlGe_data}a. At 55 kOe, the magnetization only reaches $\approx$ 0.7 $\mu_B$/Gd, far from the expected saturated moment of 7 $\mu_B$ for Gd. In their recent work, Laha et al. observe a metamagnetic transition at 62 kOe; however, the step in the magnetization is small ($\approx$ 0.2 $\mu_{\text{}B}$) and \textit{M} remains unsaturated up to at least 70 kOe. \cite{laha2025single} The magnetoresistance of GdAlGe is shown in Fig. \ref{GdAlGe_data}f, and increases in an approximately linear fashion to a maximum of 120 $\%$ at 55 kOe. Like the 2 K \textit{M}(\textit{H}) isotherm, there is no indication of saturation up to the highest measured field.

Finally, Fig. \ref{GdAlGe_data}g shows the specific heat of GdAlGe measured between 1.8--50 K. Both magnetic transitions are apparent as well-resolved peaks at \textit{T}$_{N1}$ = 34.7 K and \textit{T}$_{N2}$ = 29.8 K, consistent with the transition temperatures determined by magnetic and transport data. To estimate the magnetic contribution of the specific heat (\textit{C}$_{\text{mag}}$), we measured and subtracted the specific heat of LuAlGe, which serves as a reference for the electronic and phonon contributions to \textit{C}$_{\text{p}}$ for the heavy \textit{R}AlGe series. The data for LuAlGe is shown as a gray curve in Fig. \ref{GdAlGe_data}g and the estimated \textit{C}$_{\text{mag}}$ of GdAlGe is the green dataset. We integrated \textit{C}$_{mag}$/\textit{T} to estimate the magnetic entropy associated with the transitions. The results are shown in Fig. \ref{GdAlGe_data}h and show that approximately 90 $\%$ of the full \textit{R}ln8 is recovered by 50 K. This entropy deficiency strongly suggests that LuAlGe is only an approximate of the non-magnetic contributions to the specific heat of GdAlGe. It is also possible that some magnetic correlations persist above $T_{\text{N}}$  As such, we can only extract qualitative statements about entropy associated with other members of the RAlGe series, likely with increasing accuracy as the moment bearing R gets closer to Lu.

\begin{figure*}[t!]
    \centering
    \includegraphics[width=\linewidth]{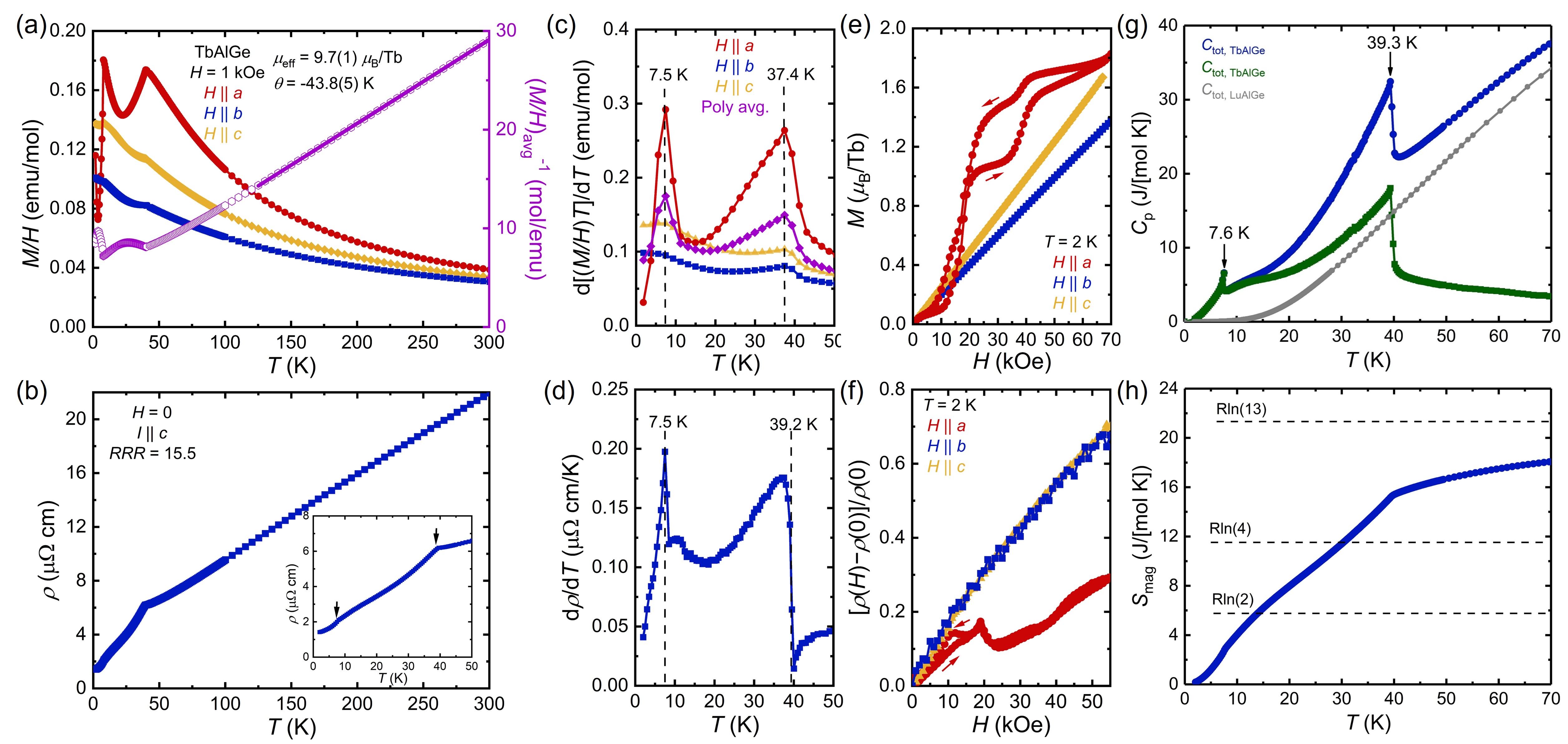}
    \caption[]{Physical properties of TbAlGe. (a) Temperature dependent magnetic susceptibility (left axis). The right axis shows the polycrystalline average of the individual inverse (\textit{M/H})$^{-1}$ curves for each orientation. The solid line shows the Curie-Weiss fit to the ($M/H$)$_{\text{avg}}^{-1}$ data (open points). (b) Temperature dependent resistivity. The inset gives a close-up view of the data between 1.8--50 K. (c) Derivative of the temperature-susceptibility product (d\textit{(M/H)T}/d\textit{T}), (d) derivative of the resistivity (d$\rho$/d\textit{T}), (e) field dependent magnetization at \textit{T} = 2 K, (f) magnetoresistance at \textit{T} = 2 K, (g) total and magnetic contributions to the heat capacity. The data for LuAlGe is also shown for reference. (h) Estimated magnetic entropy. The horizontal dashed lines show the entropy expected for different multiplets.}
    \label{TbAlGe_data}
\end{figure*}

\subsubsection{TbAlGe}

The data collected on TbAlGe single crystals are presented next in Fig. \ref{TbAlGe_data}. The magnetization is anisotropic in the paramagnetic state, with the strongest response observed when \textit{H} $\parallel$ \textit{a}. The inverse (\textit{M/H})$_{\text{avg}}^{-1}$ is linear above $\approx$ 100 K, and Curie-Weiss fits over this temperature range give $\theta_{\text{avg}}$ = -43.8(5) K, suggesting antiferromagnetic correlations, and an effective moment of 9.7(1) $\mu_B$/Tb, agreeing well with the expected 9.72 $\mu_B$ for Tb$^{3+}$.

In prior works, TbAlGe was  reported to order antiferromagnetically at \textit{T}$_{N1}$ $\approx$ 39 K and \textit{T}$_{N2}$ $\approx$ 7 K \cite{wang2021correlation,gouda2022angular,kumar2026anisotropic}, and our measurements are consistent with the earlier results. Figs. \ref{TbAlGe_data}a and \ref{TbAlGe_data}b, and the respective derivatives in \ref{TbAlGe_data}c and \ref{TbAlGe_data}d show that antiferromagnetic transitions are observed at \textit{T}$_{\textit{N1}}$ = 39 K and \textit{T}$_{\textit{N2}}$ = 7 K in our magnetic and transport data. Given the step-like drop of \textit{M/H} at 7 K when \textit{H} $\parallel$ \textit{a}, the lower transition may be first-order. The \textit{M/H} data and \textit{M}(\textit{H}) isotherms (discussed next) suggest that Tb spins are primarily aligned along the \textit{a}-axis below \textit{T}$_{N2}$. 

Fig. \ref{TbAlGe_data}e shows the field dependent magnetization of TbAlGe measured at 2 K. When \textit{H} $\parallel$ \textit{a}, \textit{M} undergoes two step-like jumps, at $\approx$ 12 kOe and 35 kOe, reaching 1.8 $\mu_B$/Tb$^{-1}$ at 70 kOe. Both transitions are hysteretic, suggesting a first order nature. At the highest fields ($\approx$ 68--70 kOe), \textit{M} appears to slightly increase, hinting that a third metamagnetic transition may occur just above 70 kOe. The possibility of another transition at higher field is consistent with the relatively low values of \textit{M} compared to the anticipated saturated moment of 9 $\mu_B$ for Tb$^{3+}$ ions; however, higher field measurements are needed to confirm this. The magnetization for \textit{H} $\parallel$ \textit{b} and \textit{H} $\parallel$ \textit{c} both show a linear field dependence with no indication of metamagnetism up to 70 kOe.  Our data is in reasonable agreement with recent work on TbAlGe single crystals reported by Gouda et al. \cite{gouda2022angular}

The magnetoresistance at 2 K is given in Fig. \ref{TbAlGe_data}f. With the field applied along \textit{a}, the resistance shows a peak at 19 kOe and a broader increase beginning at $\approx$ 40 kOe. Both features correspond reasonably well with the two metamagnetic transitions observed in the \textit{H} $\parallel$ \textit{a} magnetization isotherms. Whith the field applied along \textit{b} and \textit{c}, the magnetoresistance increases monotonically to modest values near 70 $\%$ at 55 kOe and shows no signatures of a transition, in agreement with the magnetic data for these orientations.

Both AFM transitions show strong peaks in the specific heat data (Fig. \ref{TbAlGe_data}g), and the transition temperatures \textit{T}$_{N1}$ = 39.3 K and \textit{T}$_{N2}$ = 7.6 K are in good agreement with the magnetic and transport data. The sharp, spike-like feature observed at \textit{T}$_{\text{N2}}$ may suggest the lower temperature transition is first-order, and this interpretion is consistent with the steplike drop observed in the \textit{H} $\parallel$ \textit{a} magnetic data in Fig. \ref{TbAlGe_data}a. Fig. \ref{TbAlGe_data}h shows that by 70 K, the magnetic entropy recovers $\approx$ 85$\%$ of the \textit{R}ln13 expected for \textit{J} = 6 Tb local moments. Considering the clear magnetic anisotropy visible in the \textit{M}(\textit{T}) data up to 300 K, this is consistent with some degree of CEF splitting above the magnetic ordering temperatures.

\begin{figure*}[t!]
    \centering
    \includegraphics[width=\linewidth]{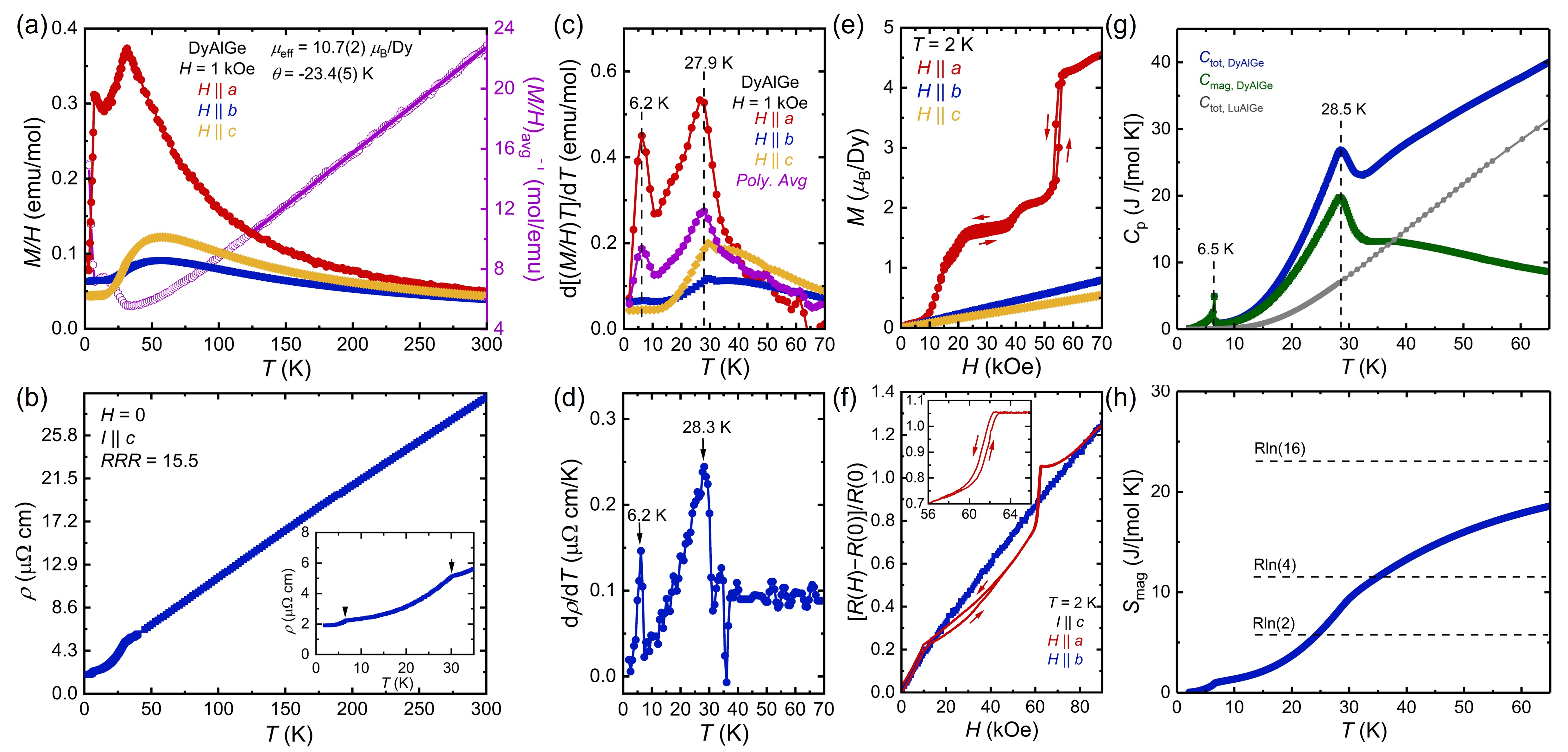}
    \caption[]{Physical properties of DyAlGe. (a) Temperature dependent magnetic susceptibility (left axis). The right axis shows the polycrystalline average of the individual inverse (\textit{M/H})$^{-1}$ curves for each orientation. The solid line shows the Curie-Weiss fit to the ($M/H$)$_{\text{avg}}^{-1}$ data (open points). (b) Temperature dependent resistivity. The inset in (b) gives a close-up of the data between 1.8--35 K with arrows marking the transitions. (c) Derivative of the temperature-susceptibility product (d\textit{(M/H)T}/d\textit{T}), (d) derivative of the resistivity (d$\rho$/d\textit{T}), (e) field dependent magnetization at \textit{T} = 2 K, (f) magnetoresistance at \textit{T} = 2 K, where the inset shows a zoomed-in view near the 60 kOe transition, (g) total and magnetic contributions to the heat capacity. The data for LuAlGe is also shown for reference. (h) Estimated magnetic entropy. The horizontal dashed lines show the entropy expected for different multiplets.}
    \label{DyAlGe_data}
\end{figure*}

\subsubsection{DyAlGe}

DyAlGe has qualitatively similar properties to TbAlGe but with stronger \textit{a}-axis anisotropy (at low temperatures) and transitions that are shifted to lower temperatures. In agreement with the data on polycrystalline samples, \cite{wang2022dynamic} our magnetic susceptibility and resistivity measurements indicate DyAlGe undergoes two transitions, at \textit{T}$_{N1}$ $\approx$ 28 K and \textit{T}$_{N2}$ $\approx$ 6.2 K, as shown in Figs. \ref{DyAlGe_data}a--\ref{DyAlGe_data}d. \textit{M/H} has a prominent peak at \textit{T}$_{N1}$ (28 K) when \textit{H} $\parallel$ \textit{a}. Broad \textit{M/H} humps with maxima at $\approx$ 50 K are observed when the field is oriented along \textit{b} and \textit{c}, likely attributable to low lying crystal field levels. This behavior shows DyAlGe to have strongly uniaxial behavior in which the moments are aligned along the \textit{a}-axis below $T_{\text{N}}$. The \textit{H} $\parallel$ \textit{a} susceptibility curve shows a second, very sharp transition at \textit{T}$_{N2}$ $\approx$ 6 K that is only weakly resolved with the field along the other principle axes but is clearly observed in the resistivity (see inset to \ref{DyAlGe_data}b). Curie-Weiss fits to the high temperature data give $\mu_{\text{eff}}$ = 10.7(2) $\mu_B$/Dy, in very good agreement with the expected 10.6 $\mu_B$ for Dy$^{3+}$ local moments. The paramagnetic temperature $\theta_{avg}$ is -23.4(5) K, consistent with the observed antiferromagnetic order. 

The field dependent magnetization of DyAlGe at 2 K is shown in Fig. \ref{DyAlGe_data}e. Like TbAlGe, rich metamagnetism is observed with the applied field along the \textit{a}-axis. The \textit{H} $\parallel$ \textit{a} data provides evidence for four metamagnetic transitions occuring below 70 kO, that occur at approximately 9 kOe, 16 kOe, 38 kOe, and 53 kOe when increasing the field. The \textit{H} $\parallel$ \textit{a} isotherm shows clear hysteresis between 14--40 kOe and for the high field transition at 52--57 kOe. For fields aligned along the \textit{b} and \textit{c} axes, the magnetization increases linearly up to 70 kOe, with a greater slope when \textit{H} $\parallel$ \textit{b}. Given that \textit{M} only reaches $\approx$ 4.5 $\mu_B$/Dy and has yet to saturate at 70 kOe, additional metamagnetic transitions are likely to be found at higher fields.  As such, DyAlGe may be like TbNi$_2$Ge$_2$, \cite{bud1999anisotropy} providing another example of a strongly uniaxial moment with a devil's staircase-like \textit{M}(\textit{H}).

The magnetoresistance is shown in Fig. \ref{DyAlGe_data}f. When the field is applied along the \textit{a}-axis, metamagnetic transitions occur at $\approx$ 10 kOe and 60 kOe. These transitions approximately align with the first and last metamagnetic transitions observed in the magnetic isotherms, and the slightly higher onset fields likely come from imperfect sample orientation. Whereas the intermediate-field metamagnetism detected in Fig. \ref{DyAlGe_data}e is not observed in the transport data, the magnetoresistance shows a broad hysteretic region over approximately the same fields, between 10--45 kOe. The magnetoresistance increases monotonically with field when \textit{H} $\parallel$ \textit{b}, consistent with the magnetic data, and reaches modest maximum values of 120 $\%$ at 90 kOe.

Fig. \ref{DyAlGe_data}g shows that both transitions are observed in the specific heat, and the transition temperatures agree well with the magnetic and transport results. While in part accounted for by an imperfect non-magnetic background subtraction (LuAlGe), there is substantial magnetic entropy remaining above \textit{T}$_N$, and the peak associated with magnetic order at 28.5 K appears superimposed on a relatively broad Schottky-like maxima in \textit{C}$_{\text{mag}}$. This is consistent with \textit{H} $\parallel$ \textit{b} and \textit{H} $\parallel$ \textit{c} magnetization data, and is associated with thermal population of the crystal field levels. Given that an analogous feature is not observed in TbAlGe, which also shows strongly uniaxial behavior in the paramagnetic state (see Fig. \ref{X-1anis} in the Appendix) but much weaker anisotropy at 2 K, it is likely that the ground state CEF multiplet is more energetically isolated from the next lowest energy states in DyAlGe than in TbAlGe. Inelastic neutron data would be needed to confirm this picture. By 70 K, the integrated magnetic entropy reaches nearly 80 $\%$ the full \textit{R}ln16 expected for trivalent Dy moments.

\begin{figure*}[t!]
    \centering
    \includegraphics[width=\linewidth]{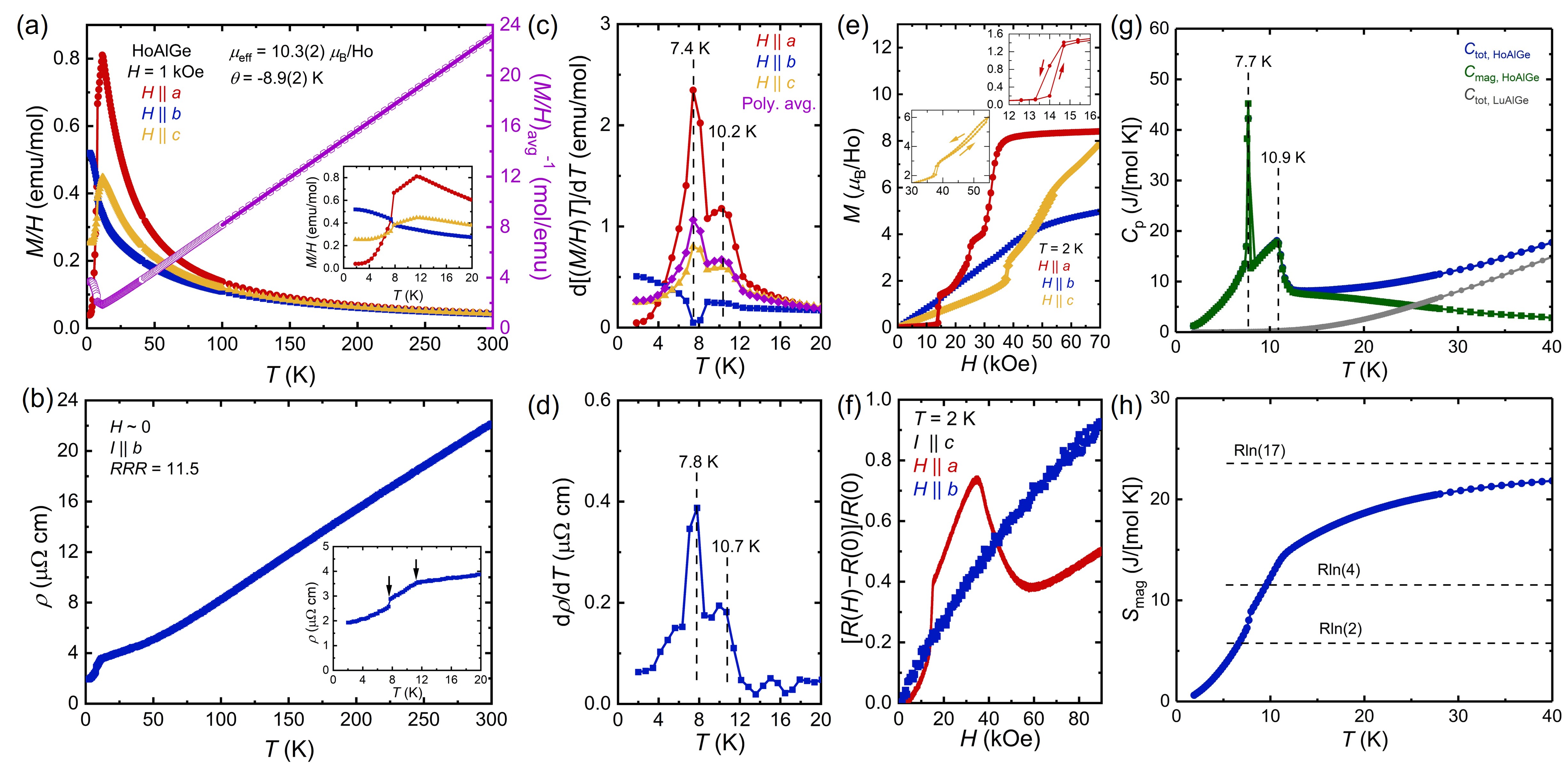}
    \caption[]{Physical properties of HoAlGe. (a) Temperature dependent magnetic susceptibility (left axis). The right axis shows the polycrystalline average of the individual inverse (\textit{M/H})$^{-1}$ curves for each orientation.  The solid line shows the Curie-Weiss fit to the ($M/H$)$_{\text{avg}}^{-1}$ data (open points). (b) Temperature dependent resistivity. The insets in (a) and (b) give zoomed view of the data between 1.8--20 K. (c) Derivative of the temperature-susceptibility product (d\textit{(M/H)T}/d\textit{T}), (d) derivative of the resistivity (d$\rho$/d\textit{T}), (e) field dependent magnetization at \textit{T} = 2 K with insets showing a close-up view near the metamagnetic transitions, (f) magnetoresistance at \textit{T} = 2 K, (g) total and magnetic contributions to the heat capacity. The data for LuAlGe is also shown for reference. (h) Estimated magnetic entropy. The horizontal dashed lines show the entropy expected for different multiplets.}
    \label{HoAlGe_data}
\end{figure*}

\subsubsection{HoAlGe}

Fig. \ref{HoAlGe_data} shows the data collected on HoAlGe single crystals. The results are again qualitatively similar to those of TbAlGe and DyAlGe, with \textit{T}$_{\text{N1}}$ pushed to lower temperature as the \textit{R} atom changes from Tb to Ho. The magnetic susceptibility (Fig. \ref{HoAlGe_data}a) shows two antiferromagnetic transitions, at \textit{T}$_{N1}$ = 10.2 K and \textit{T}$_{N2}$ = 7.4 K based on d\textit{(M/H)T}/d\textit{T} (Fig. \ref{HoAlGe_data}c). The resistivity, shown in Fig. \ref{HoAlGe_data}b shows good metallic behavior, a \textit{RRR} of 11.5, and two transitions in excellent agreement with the magnetic results (see d$\rho$/d\textit{T} in Fig. \ref{HoAlGe_data}d). These results are again roughly consistent with 11 K and 8 K antiferromagnetic transition temperatures previously found in polycrystalline HoAlGe. \cite{wang2022dynamic} Fitting the high temperature susceptibility to the Curie-Weiss law gives $\theta$ = -8.9(2) K, consistent with antiferromagnetic interactions. The estimated effective moment is 10.3(2) $\mu_B$/Ho, matching the expected value of 10.3 for trivalent Ho local moments. 

The field dependent magnetization at 2 K is shown in Fig. \ref{HoAlGe_data}e. When \textit{H} $\parallel$ \textit{a}, Fig. \ref{HoAlGe_data}e shows three metamagetic transitions at approximately 14 kOe, 25 kOe, and 32.5 kOe. Here the critical fields were determined by maxima in d\textit{M}/d\textit{H}, for the increasing field-sweep. The upper inset in Fig. \ref{HoAlGe_data}e shows clear hysteresis between $\approx$ 12--25 kOe, whereas the higher field transition at 30 kOe is not hysteretic. Above the transitions, the magnetization saturates at $\approx$ 8.5 $\mu_B$ at 70 kOe. When \textit{H} $\parallel$ \textit{c}, the magnetization first increases linearly at low fields and then undergoes two metamagnetic transitions at approximately 38 kOe and 48 kOe. Both transitions display measurable hysteresis upon raising and lowering the field, and above the second transition, the magnetization continues to increase with field, reaching nearly 8 $\mu_B$ at 70 kOe. As this is smaller than the expected 10 $\mu_B$ expected for Ho$^{3+}$, we anticipate additional metamagnetism at yet higher fields. Finally, when \textit{H} $\parallel$ \textit{b}, the magnetization increases linearly to $\approx$ 4 $\mu_B$ at 40 kOe, where the curve begins to saturate towards 5 $\mu_B$ as \textit{H} approaches 70 kOe. The presence of metamagnetism in multiple field orientations is consistent with the temperature dependent behavior, indicating that while still anisotropic, the magnetism of HoAlGe is substantially less anisotropic than in DyAlGe. The magnetorestance is shown in Fig. \ref{HoAlGe_data}f and undergoes a step-like jump at 13 kOe when \textit{H} $\parallel$ \textit{a}. Above the metamagnetic transition, the resistance increases nearly linearly with field until 36 kOe, where it drops rapidly, finally reaching a minimum at 58 kOe, above which it increases with field and is not saturated by 90 kOe. Like with DyAlGe, the sharp changes in behavior shown in the magnetoresistance for HoAlGe correspond reasonably well with the metamagnetism found in the \textit{M}(\textit{H}) isotherms.  

Fig. \ref{HoAlGe_data}g shows the heat capacity of HoAlGe. Both transitions are evident at 7.7 K and 10.9 K, in good agreement with magnetic and transport results. The lower, 7.7 K, transition produces a very narrow spike in \textit{C}$_{\text{p}}$, indicating a first order nature. This is consistent with the sharp, step-like, features observed in the magnetic and resistivity datasets at \textit{T}$_\text{N2}$ (see insets to Figs. \ref{HoAlGe_data}a and \ref{HoAlGe_data}b). The magnetic entropy, shown in Fig. \ref{HoAlGe_data}h, recovers nearly the full \textit{R}ln17 expected for free $J$ = 9 Ho$^{3+}$ ions by 40 K, suggesting HoAlGe has smaller CEF splitting compared to the $R$ = Tb and Dy members which is consistent with the weaker magnetic anisotropy observed for HoAlGe. 

\begin{figure*}[t!]
    \centering
    \includegraphics[width=\linewidth]{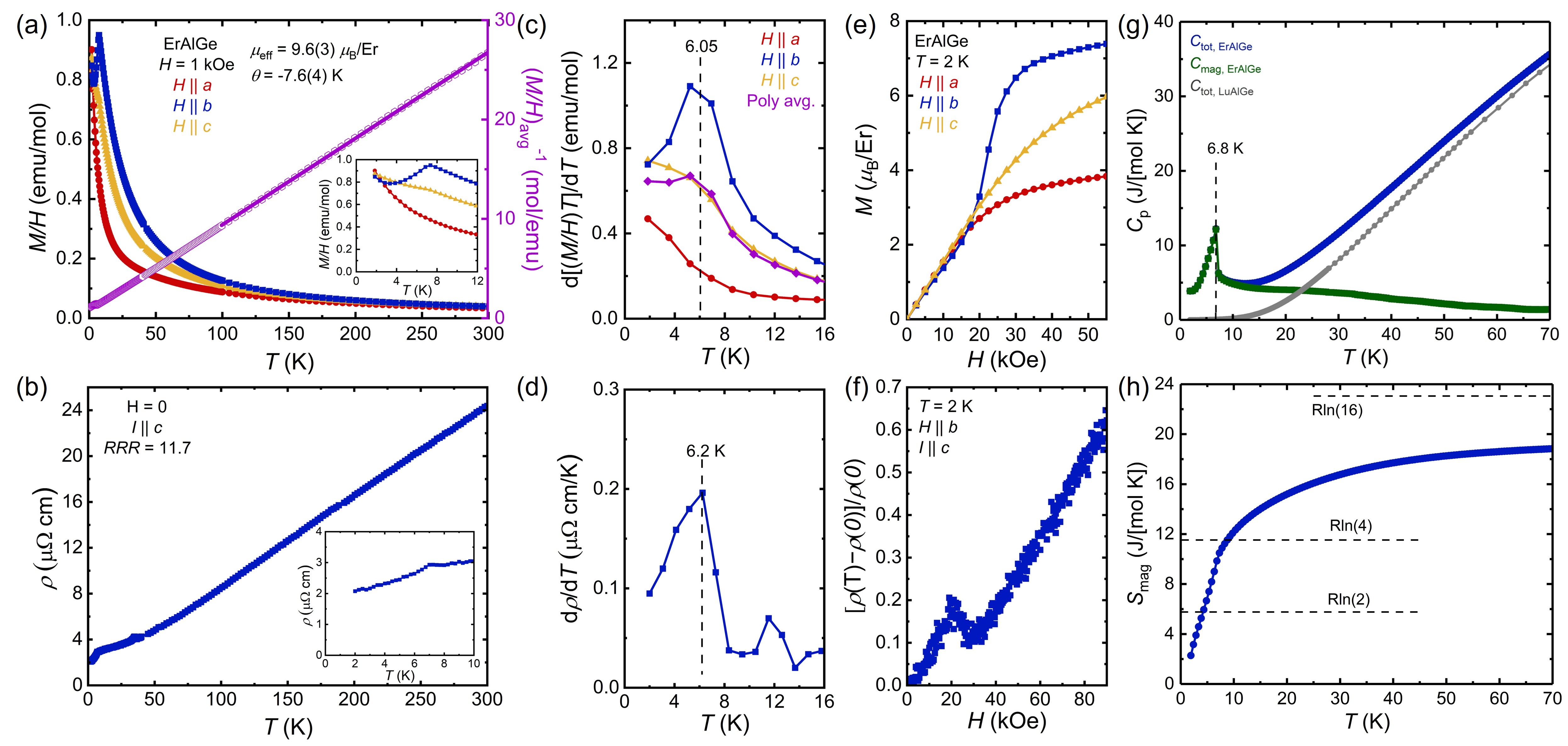}
    \caption[]{Physical properties of ErAlGe. (a) Temperature dependent magnetic susceptibility (left axis). The right axis shows the average of the individual inverse (\textit{M/H})$^{-1}$ curves for each orientation. The solid line shows the Curie-Weiss fit to the ($M/H$)$_{\text{avg}}^{-1}$ data (open points). (b) Temperature dependent resistivity. The insets in (a) and (b) give a zoomed view of the data at low temperature. (c) Derivative of the temperature-susceptibility product (d\textit{(M/H)T}/d\textit{T}), (d) derivative of the resistivity (d$\rho$/d\textit{T}), (e) field dependent magnetization at \textit{T} = 2 K, (f) magnetoresistance at \textit{T} = 2 K, (g) total and magnetic contributions to the heat capacity. The data for LuAlGe is also shown for reference. (h) Estimated magnetic entropy. The horizontal dashed lines show the entropy expected for different multiplets..}
    \label{ErAlGe_data}
\end{figure*}

\subsubsection{ErAlGe}

The magnetic behavior of the \textit{R}AlGe series substantially changes when crossing from \textit{R} = Ho to Er. The temperature dependent magnetization and resistivity, and their respective derivatives shown in Figs. \ref{ErAlGe_data}a-\ref{ErAlGe_data}d indicate that the Er moments order antiferromagnetically at \textit{T}$_\text{N}$ = 6 K, with no additional transitions observed at lower temperature. Fig. \ref{ErAlGe_data}a and its inset show that whereas \textit{M/H} reaches a maximum at the Néel temperature when \textit{H} $\parallel$ \textit{b}, the transition is only weakly observed when the field is applied along \textit{a}- or \textit{c}-axes. Unlike samples with \textit{R} = Tb-Ho, which likely have local moments oriented primarily along the \textit{a}-axis, for \textit{R} = Er, the behavior above and below \textit{T}$_N$ suggests the local moments have an easy \textit{b}-axis. It is very likely that the change in anisotropy observed here in the orthorhombic \textit{R}AlGe family reflects a change in sign of the leading terms in the crystal field Hamiltonian, which are normally expected to switch between between $R$ = Ho and $R$ = Er in an isostructural series owing to the change in sign of the second-order Steven's factor \cite{HUTCHINGS1964227}. This is discussed in more detail in the discussion section.

The right axis of Fig. \ref{ErAlGe_data}a shows the inverse of \textit{M/H}$_\text{avg}$ of ErAlGe. The Curie-Weiss fits suggest an average effective moment of 9.6(3) $\mu_B$/Er, in excellent agreement with the expected value of 9.59 $\mu_B$ for Er$^{3+}$ local moments. The average Weiss temperature is -7.6(4) K, suggesting antiferromagnetic interactions above \textit{T}$_N$, which is also in agreement with the experimental observations. 

Fig. \ref{ErAlGe_data}e shows the field dependent magnetization at 2 K. \textit{M} is isotropic until $\approx$20 kOe, upon which \textit{M} rapidly jumps from $\approx$ 3 $\mu_B$/Er to 7 $\mu_B$/Er between 20--30 kOe when \textit{H} $\parallel$ \textit{b}. Above 20 kOe, when \textit{H} $\parallel$ \textit{a}, \textit{M} appears to be saturating towards $\approx$ 4 $\mu_B$/Er, and when \textit{H} $\parallel$ \textit{c}, \textit{M} continues to gradually increase, approaching 7.5 $\mu_B$/Er at 55 kOe, which is lower than the 9 $\mu_B$ expected for Er$^{3+}$ local moments. The isotherm shows no hysteresis when measuring with increasing and decreasing fields, differing from the metamagnetism observed for Tb--Ho based samples. The \textit{H} $\parallel$ \textit{b} magnetoresitance, shown in Fig. \ref{ErAlGe_data}f, has a prominent peak at $\approx$ 20 kOe, in agreement with the metamagnetic transition, and otherwise increases with field up to $\approx$ 60 $\%$ at 90 kOe.

The heat capacity of ErAlGe is shown in Fig. \ref{ErAlGe_data}g. A lambda-like peak is observed at 6.8 K, in reasonable agreement with the Néel temperature inferred from the magnetic and transport data. Below the transition, the heat capacity does not approach zero and is nearly constant on cooling at the lowest temperatures. This behavior indicates there is remaining entropy at our base temperature of 1.9 K and suggests that another magnetic transition would likely occur with further cooling. The magnetic entropy, shown in Fig. \ref{ErAlGe_data}h recovers nearly \textit{R}ln4 at the transition. Given a formal CEF quartet is impossible in orthorhombic symmetry, this may indicate the presence of a low-energy pseudoquartet, composed of two closely spaced doublets. By 70 K, the magnetic entropy recovers approximately 80 $\%$ the \textit{R}ln16 expected for trivalent Er moments. Considering that the low temperature specific heat is not trending towards the origin, the spline extrapolation to $C_{\text{p}}$ = 0 at $T$ = 0 that we used to estimate the specific heat below our lowest temperature data point (see experimental details section) is unreliable in ErAlGe. Because the true temperature dependence of $C_{\text{p}}$ was not measured below 1.9 K, the integration shown in Fig. \ref{ErAlGe_data}h likely results in an underestimate of the true magnetic entropy.

\begin{figure*}[t!]
    \centering
    \includegraphics[width=\linewidth]{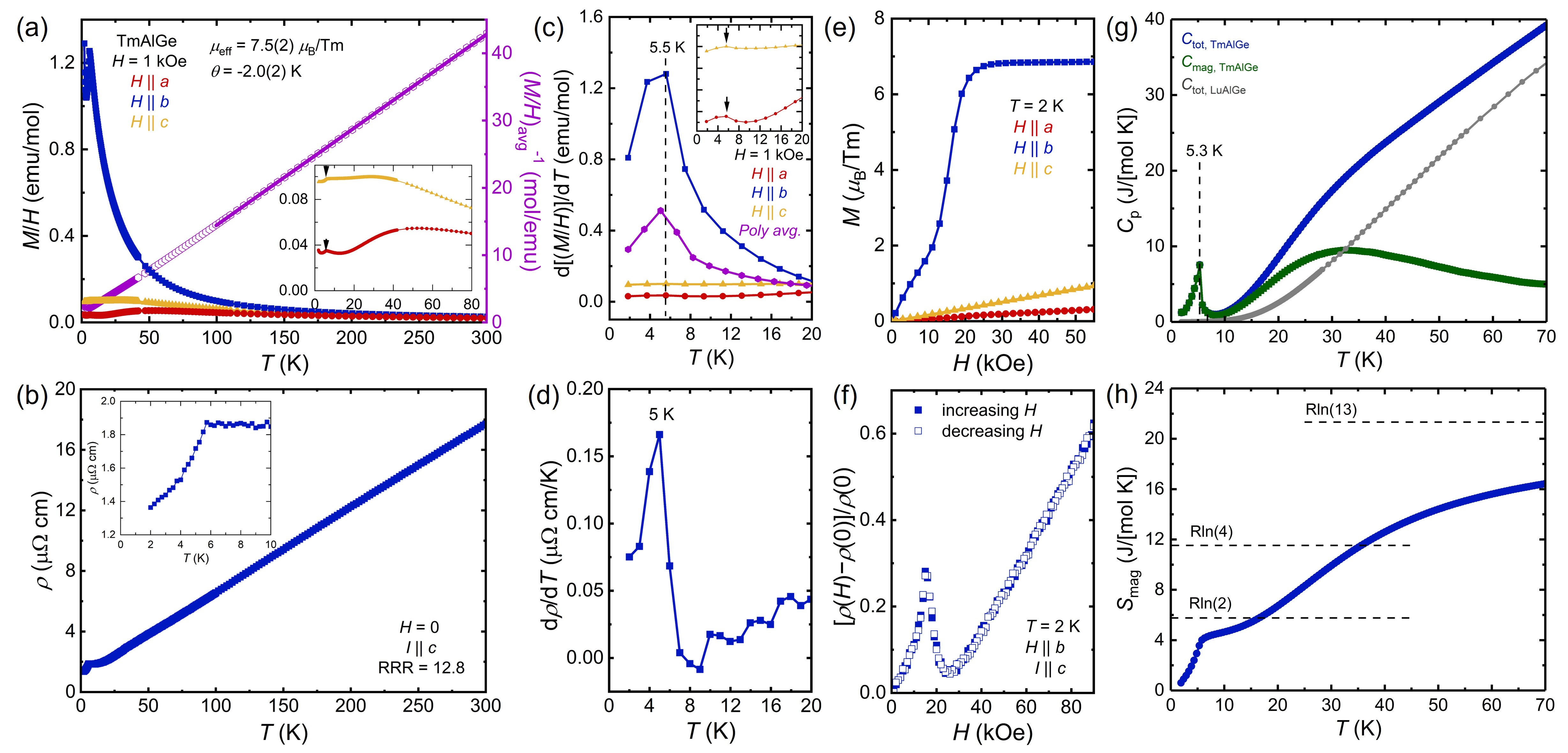}
    \caption[]{Physical properties of TmAlGe. (a) Temperature dependent magnetic susceptibility (left axis). The right axis shows the average of the individual inverse (\textit{M/H})$^{-1}$ curves for each orientation. The solid line shows the Curie-Weiss fit to the ($M/H$)$_{\text{avg}}^{-1}$ data (open points). (b) Temperature dependent resistivity. (c) Derivative of the temperature-susceptibility product (d\textit{(M/H)T}/d\textit{T}). The insets in (a)--(c) give zoomed views of the data at low temperatures.  (d) Derivative of the resistivity (d$\rho$/d\textit{T}), (e) field dependent magnetization at \textit{T} = 2 K, (f) magnetoresistance at \textit{T} = 2 K, (g) total and magnetic contributions to the heat capacity. The data for LuAlGe is also shown for reference. (h) Estimated magnetic entropy. The horizontal dashed lines show the entropy expected for different multiplets.}
    \label{TmAlGe_data}
\end{figure*}

\subsubsection{TmAlGe}

The physical properties of TmAlGe are similar to those of ErAlGe but with much stronger \textit{b}-axis anisotropy. Like ErAlGe, the magnetization (shown in Fig. \ref{TmAlGe_data}a) shows a prominent peak at \textit{T}$_N$ = 5 K when \textit{H} $\parallel$ \textit{b}. The temperature dependence is considerably weaker for samples oriented with the field applied along \textit{a} and \textit{c}. In both orientations, a small local \textit{M/H} maxima is observed at 5 K followed by another, relatively broad, maxima between $\approx$ 10-70 K (shown in inset to Fig. \ref{TmAlGe_data}a). The latter is likely attributable to relatively low-lying crystal field multiplets. Fitting \textit{M/H}$^{-1}_{\text{avg}}$ to the Curie-Weiss law gives $\theta_{\text{avg}}$ = -2.0(2) and $\mu_{\text{eff}}$ = 7.5(2) $\mu_B$/Tm, in good agreement with the expectation of 7.56 $\mu_B$/Tm. 

The resistivity for TmAlGe is shown in Fig. \ref{TmAlGe_data}b. Like the other \textit{R}AlGe materials, ErAlGe shows metallic behavior and a moderate \textit{RRR} of 12.8. Loss of spin disorder scattering is evident at \textit{T}$_N$ $\approx$ 5 K (see d$\rho$/\textit{T} in Fig. \ref{TmAlGe_data}d), in good agreement with the single transition found at 5 K in the magnetic data. 

The field dependent magnetization for TmAlGe is presented in Fig. \ref{TmAlGe_data}e. When \textit{H} $\parallel$ \textit{b}, a clear metamagnetic transition is observed at $\approx$ 12 kOe, where the magnetization quickly jumps from 2 to 7 $\mu_B$/Tm. Like ErAlGe, the metamagnetic transition does not have observable hysteresis when raising or lowing the field; however, unlike ErAlGe, TmAlGe shows substantial \textit{M}(\textit{H}) anisotropy at all fields, whereas the isotherms for ErAlGe are isotropic below its metamagnetic transition. Above 25 kOe, the magnetization for the \textit{H} $\parallel$ \textit{b} orientation appears saturated at 6.9 $\mu_{\textit{B}}$ up to 55 kOe, in reasonable agreement with the expected 7 $\mu_B$ for Tm ions. No metamagnetism is evident in the other two orientations. Instead, \textit{M} increases linearly over the measured field range when \textit{H} $\parallel$ \textit{c}, reaching 1 $\mu_B$/Tm at 55 kOe. \textit{M} increases weakly with the field when oriented with \textit{H} $\parallel$ \textit{a}, only achieving 0.3 $\mu_B$/Tm at 55 kOe.

Fig. \ref{TmAlGe_data}f gives the magnetoresistance of TmAlGe when \textit{I} $\parallel$ \textit{c} and \textit{H} $\parallel$ \textit{b}. Like ErAlGe, the magnetoresistance first increases with field, but falls sharply at 15 kOe, corresponding to the metamagnetic transition observed for \textit{H} $\parallel$ \textit{b} in the field dependent magnetization data (Fig. \ref{TmAlGe_data}e). After the peak, the resistance increases nearly linearly up to 60$\%$ at 90 kOe.

Finally, Fig. \ref{TmAlGe_data}g shows the heat capacity of TmAlGe. A lambda peak is observed at 5.3 K, consistent with the Néel temperature infered from magnetic and transport data. Like ErAlGe, the specific heat does not trend towards zero at our lowest temperature, which may point towards additional magnetic transitions at lower temperatures. The magnetic heat capacity furthermore shows a broad Schottky-like anomaly between $\approx$ 10-70 K, likely reflecting thermal population of the crystal field levels. This is consistent with the behavior of the magnetization when the field is applied along the \textit{a}- and \textit{c}-axis and in-line with the much stronger uniaxial anisotropy oberved for TmAlGe than ErAlGe. The magnetic entropy is shown in Fig. \ref{TmAlGe_data}h and recovers nearly 75 $\%$ \textit{R}ln2 at the 5.3 K transition, which may suggest a ground state doublet. Since Tm is not a Kramer's ion, such a doublet is not guaranteed. The integrated entropy recovers over 75 $\%$ of the full \textit{R}ln13 by 70 K. As was the case for ErAlGe, the magnetic entropy in Fig. \ref{TmAlGe_data}h is likely underestimated due to the small but apparent remaining entropy at 1.9 K, which is not accounted for by our extrapolation of the data to the origin and therefore neglected in our integration.

\begin{table*}[htbp]
    \centering
    \caption{Summary of key results for the moment-bearing members of the \textit{R}AlGe family (R = Gd-Tm). Lattice parameters, antiferromagnetic transition temperatures, effective magnetic moments, averaged Weiss temperatures, and de Gennes factor (dG = $(g-J)^{2}J(J+1)$). The listed transition temperatures are inferred from the specific heat measurements. We also include references to the magnetic ordering temperatures reported in previous studies of $R$AlGe compounds.}
    \resizebox{\linewidth}{!}{\begin{tabular}{cccccccccc}
    \toprule
        RAlGe & \textit{a} (\AA)  & \textit{b} (\AA)  & \textit{c} (\AA)  & \textit{V} (\AA$^3$)  & \textit{T}$_{\text{N1}}$ (K)  & \textit{T}$_{\text{N2}}$ (K) & $\mu_{\text{eff}}$ ($\mu_{\text{B}}$/\textit{R})  & $\theta_{\text{avg}}$ (K) & dG \\
    \midrule
        GdAlGe & 4.0626(1) & 10.52(2)  & 5.803(1)  & 248.05(1)  & 34.7, 35 \cite{laha2025single}  & 29.8, 30 \cite{laha2025single}  & 7.9(1)  & -66(1)  & 15.75 \\
        TbAlGe & 4.0467(1)  & 10.45(2)  & 5.772(1)  & 244.00(1)  & 39.3, 39 \cite{wang2021correlation}, 38 \cite{gouda2022angular}, 40 \cite{kumar2026anisotropic}  & 7.6, 7 \cite{wang2021correlation}, 7.6 \cite{gouda2022angular}, 8 \cite{kumar2026anisotropic}  & 9.7(1)  & -43.8(5)  & 10.5 \\
        DyAlGe & 4.0376(1)  & 10.40(2)  & 5.749(1)  & 241.39(1)  & 28.5, 30 \cite{wang2022dynamic}  & 6.5  & 10.7(2)  & -23.4(5)  & 7.08 \\
        HoAlGe & 4.0306(1) & 10.35(2)  & 5.729(1)  & 239.08(1)  & 10.9, 11 \cite{wang2022dynamic}  & 7.7  & 10.3(2)  & -8.9(2)  & 4.5 \\
        ErAlGe & 4.0222(1)  & 10.31(2)  & 5.708(1)  & 236.64(1)  & 6.8, 7 \cite{wang2021correlation}  & -  & 9.6(3)  & -7.6(4) & 2.55 \\
        TmAlGe & 4.0136(1)  & 10.27(2)  & 5.688(1)  & 234.53(1)  & 5.3  & -  & 7.5(2)  & -2.0(2)  & 1.16 \\
        \bottomrule
    \end{tabular}}
    \label{SummaryTable}
\end{table*}

\section{Discussion}

\begin{figure}[b!]
    \centering
    \includegraphics[width=\linewidth]{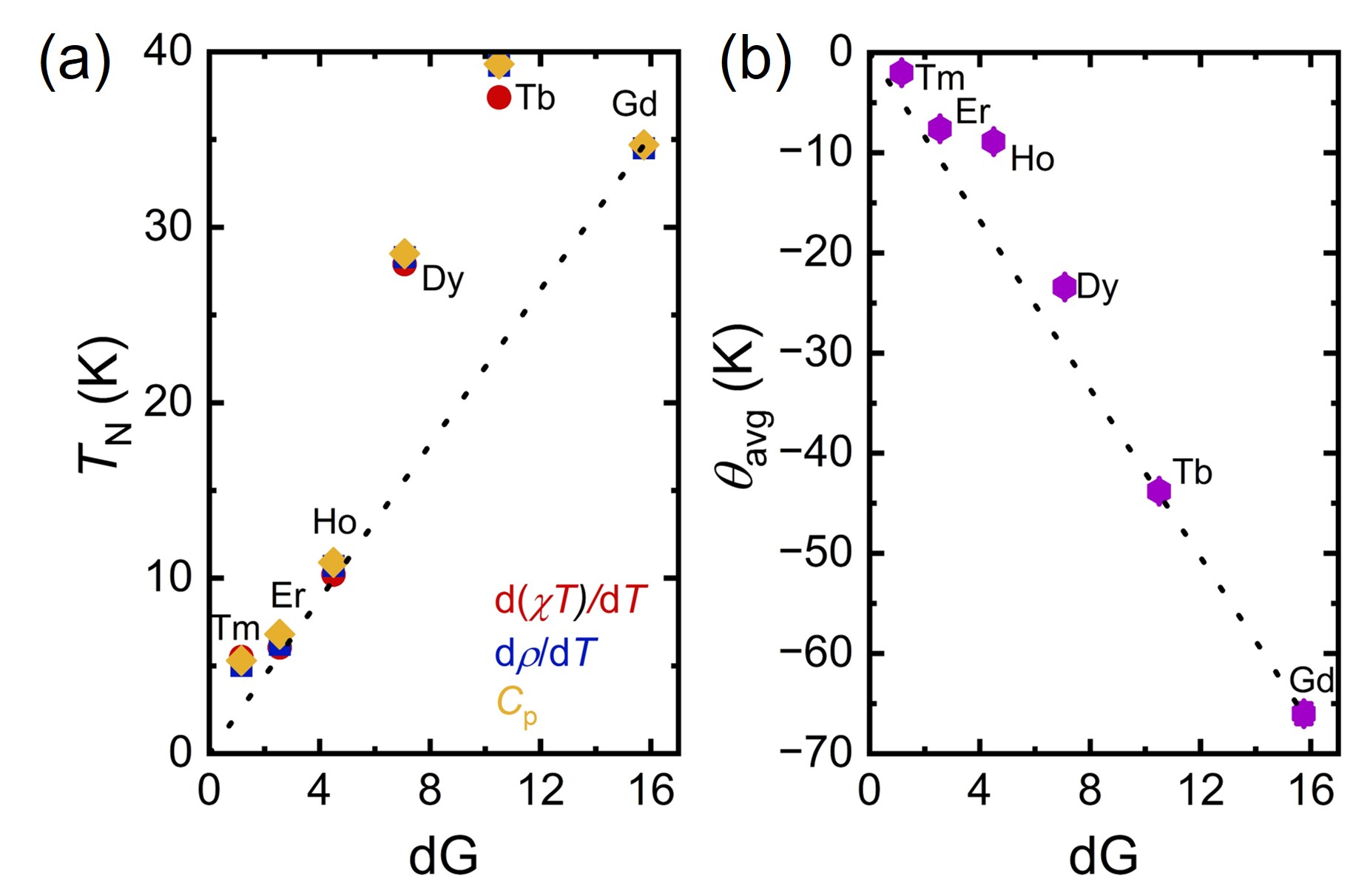}
    \caption[]{(a) Néel temperature as a function of the rare-earth ion de Gennes factor (dG) for the \textit{R}AlGe. (b) Weiss temperature ($\theta_{\text{avg}}$), obtained from polycrystalline averaged \textit{M/H}, as a function of the rare-earth ion de Gennes factor for the \textit{R}AlGe. The dashed lines show the values of $T_N$ and $\theta_{\text{avg}}$ expected for ideal de Gennes scaling.}
    \label{dG}
\end{figure}

\subsection{de Gennes scaling and magnetic anisotropy}

The key results from our measurements are outlined in Table 1, which contains the lattice parameters, magnetic transition temperatures, effective (paramagnetic) magnetic moment, Weiss temperatures, and de Gennes factor ((dG = $(g-J)^{2}J(J+1)$), with \textit{g} the Lande \textit{g} factor and \textit{J} the total angular momentum for the magnetic moment-bearing members of the \textit{R}AlGe family. At sufficiently high temperatures ($T >$ 100 K), the moment bearing $R$AlGe compounds ($R$ = Gd-Tm) show paramagnetic Curie-Weiss like behavior with estimated effective moments that agree well the the expected free-ion values. On cooling, the moment bearing members all show long-range antiferromagnetic order. When \textit{R} = Tb--Ho, the samples have two transitions in which the moments are aligned (at least primarily) along the \textit{a}-axis, with strongest anisotropy observed for DyAlGe. The first, higher temperature transition, decreases from 39 K in TbAlGe to 10 K in HoAlGe, whereas the second, lower temperature, transition occurs near $\approx$7 K in each. Based on the step-like features in the magnetization and sharp peaks observed in the specific heat, the lower transition in Tm-, Dy-, and Ho-based members is likely first-order in nature. The magnetic behavior changes for \textit{R} beyond Ho. Unlike the \textit{R} = Gd--Ho members, ErAlGe and TmAlGe each have only a single magnetic transition above 1.8 K and show \textit{b}-axis anisotropy that is considerably stronger in the case of TmAlGe. We note however that a second, lower temperature transition may also be found in these samples below 1.8 K, and that this notion is roughly supported by the small, but finite, entropy implied by the behavior of the specific heat at lowest temperatures.

If the RKKY interactions mediates the coupling between the rare-earth based local moments through the conduction electrons, then in a mean field approximation, both the magnetic ordering and paramagnetic Weiss temperature should be proportional to the de Gennes factor. As shown in Fig. \ref{dG} and Table \ref{SummaryTable}, we find relatively poor, non-linear and non-monotonic scaling between \textit{T}$_{N1}$ and \textit{dG}. Such poor de Gennes scaling of $T_N$ is likely associated with the crystal field, which constrains the moments to the $a$-axis for $R$ = Tb--Ho members and to the $b$-axis for $R$ = Er, Tm. 

Strongly uniaxial behavior produces highly anisotropic interactions between magnetic moments, conditions under which de Gennes scaling is expected to be violated \cite{noakes1982effect}. In the $R$AlGe series, this is supported by two observations. First, the deviation from the expected de Gennes scaling is greatest for $R$ = Tb, Dy, and Tm, which differ from the "ideal" values by a factor of 1.6, 1.8, and 2, respectively, whereas the $T_N$ values of HoAlGe and ErAlGe essentially fall on the expected line. If we take the magnetic anisotropy as an indicator of the crystal field splitting, the relative deviation from the values predicted for perfect de Gennes scaling is strongest for the $R$AlGe members with the greatest anisotropy (see Figs. \ref{TbAlGe_data}-\ref{TmAlGe_data} and Fig. \ref{X-1anis} in the Appendix). Second, unlike the ordering temperatures, Fig. \ref{dG}b shows that the polycrystalline average Weiss temperatures scales well with the de Gennes factor. Whereas the anisotropic Weiss temperatures ($\theta_a$, $\theta_b$, and $\theta_c$) have contributions from both the molecular field (exchange) and the crystal field, the crystal field effects should be canceled by averaging such that $\theta_{avg}$, obtained from fits to $(M/H)_{\text{avg}}^{-1}$, should reflect only the sign and magnitude of the exchange \cite{dunlap1983crystalline}. The substantially improved linearity between dG and $\theta_{avg}$ support RKKY mediated exchange in the $R$AlGe series, where the deviations in the de Gennes scaling of $T_N$ are the consequence of the crystal field restricting magnetic moments to $a$- or $b$-crystallographic axes for $R$ = Tb-Ho and $R$ = Er, Tm respectively.

The change from having an easy $a$- to $b$-axis that is observed on crossing from HoAlGe to ErAlGe is also an expected consequence of the crystal field. This can be demonstrated by considering the general crystal field Hamiltonian

\begin{equation}
    H_{CEF} = \sum_{n,m}B_n^mO_n^m
\end{equation}

\noindent where $B_n^m$ are the crystal field parameters and $O_n^m$ are the Steven's operators. In the $Cmcm$ $R$AlGe materials, the rare-earth atoms occupy a relatively low symmetry environment with $m2m$ point symmetry, producing a full CEF Hamiltonian in which all nine even terms from $B_0^2$ to $B_6^6$ are allowed. Despite the complexity, if we assume the leading terms are dominant, then the sign of $B_2^0$ will dictate whether the system shows an easy $z$-axis or $xy$-plane, and a finite $B_2^2$ will further break the rotational symmetry within the $xy$-plane. Importantly, the second-order terms are expected to change sign between $R$ = Ho and $R$ = Er, as can be seen by the general expression for the crystal field parameters \cite{scheie2021pycrystalfield}

\begin{equation}
    B_n^m = A_n^m \langle r^n \rangle \theta_n  
\end{equation}

\noindent where $A_n^m$ are related to the ligand environment, $\langle r^n \rangle$ is the expectation value of the radial wavefunction, and $\theta_n$ are the Steven's multiplicative factors \cite{stevens1952matrix,HUTCHINGS1964227}. Because the point-symmetry and ligand environment are preserved across the $R$AlGe series, the sign of $B_2^0$ is generically expected to change between HoAlGe and ErAlGe owing to the fact that the second-order Stevens parameter, $\theta_2 = \alpha_J$, switches from negative values for $R$ = Tb--Ho to positive for $R$ = Ho, Er. \cite{HUTCHINGS1964227} Consequentially, the magnetic easy-axis of the RAlGe compounds is anticipated to change between HoAlGe and ErAlGe, as is commonly observed in many rare earth-based series of materials. \cite{hashimoto1979magnetic,bud1999anisotropy,morosan2004thermodynamic,PhysRevB.7.3226,zhang2022electronic,lee2022anisotropic}

As outlined in the Appendix, we attempted to make these statements more quantitative by using a simplified crystal field model, retaining only second-order terms in the CEF Hamiltonian \cite{wang1971crystal,shohata1977magnetic,hashimoto1979magnetic}, to estimate $B_2^0$ and $B_2^2$ from the anisotropic Weiss temperatures. Whereas this approach gives a rather poor estimate of the absolute value of the CEF parameters (i.e. a wide range of possible values depending on how the parameters are estimated), we indeed observe a change in sign of both $B_2^0$ and $B_2^2$ from positive values for $R$ = Tb--Ho to negative values for $R$ = Er, Tm. Here, the $z$-axis of the crystal field will correspond to the crystallographic $b$-axis of the $R$AlGe series, which is the primary 2-fold rotation axis in the $Cmcm$ space group. From this reference frame, the $x$- and $y$- crystal field axes will align with the crystallographic $c$- and $a$-directions. As discussed explicitly in the Appendix, the evolution of the estimated $B_2^0$ and $B_2^2$ is consistent with the change from samples having an easy $a$-axis to easy $b$-axis on crossing from HoAlGe to ErAlGe. Because the absolute values of the CEF parameters are poorly determined by our simplified approach, we do not attempt to propose explicit CEF splitting schemes for the $R$AlGe compounds; however, in each case, diagonalizing the approximate CEF Hamiltonians indicates that the scale of the CEF splitting is largest for DyAlGe and TmAlGe, on the order of 150 K, and smallest ($\approx$ 50 K) in HoAlGe and ErAlGe, results that are at least qualitatively consistent with the strength of the observed magnetic ansisotropy in the $R$AlGe series. Finally, we note that point charge crystal field calculations, implemented in PyCrystalField \cite{scheie2021pycrystalfield}, give qualitatively similar trends; however, the absolute values of the crystal field parameters and resulting eigenvalues are extremely sensitive to the charges assigned to the ligand atoms and therefor provide little quantitative guidance in determining appropriate crystal field schemes for the $R$AlGe. Ultimately inelastic neutron scattering data is needed to quantitatively determine and model the crystal field splitting.

\begin{figure*}[t!]
    \centering
    \includegraphics[width=\linewidth]{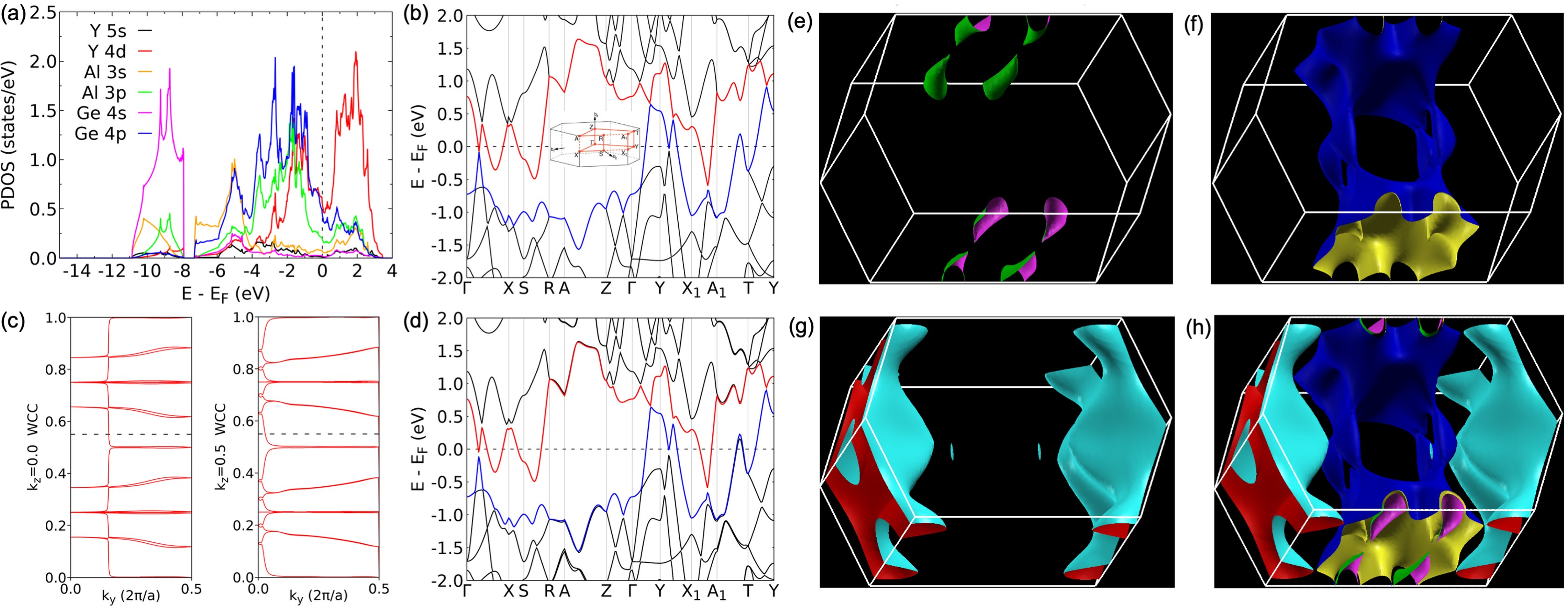}
    \caption[]{Electronic band structure of YAlGe. (a) Projected density of states (PDOS) on atomic orbitals. (b) Band structure without spin-orbit coupling (SOC). The inset shows the bulk Brillouin zone (BZ) and high-symmetry $k$-points. (c) Wilson loop with Wannier charge centers (WCC) on $k_{\text{z}}$ = 0.0 and $k_{\text{z}}$ = 0.5 planes for (d) band structure with SOC. In (b) and (d), the highest valence (lowest conduction) band is colored in blue (red). (e)--(g) show the DFT calculated Fermi surface (FS) of the three bands crossing the Fermi level. The FS corresponds to the (e) next-valence, (f) highest-valence, and (c) lowest conduction bands, respectively. (h) is the combined FS from the three bands in (e-g).}
    \label{BandStructure}
\end{figure*}

\subsection{Electronic band structure}

Given the topological physics found in the tetragonal, light \textit{R}-based, \textit{R}AlGe compounds, it is worth considering whether the orthorhombic, heavy \textit{R} members presented here may also support topological electronic states. This is further motivated by the observation of quantum oscillations in YAlGe (see Fig. \ref{YAlGe_LuAlGe}), which show that the non-magnetic band structure contains small pockets. To explore this question, we used DFT to calculate the electronic band structure of non-magnetic YAlGe. Fig. \ref{BandStructure}a plots the projected density of states (PDOS) on atomic orbitals for YAlGe. While Ge 4$s$ orbitals dominate the lower energy states from -11.0 to -8.0 eV, other orbitals hybridize strongly from -7.5 to +3.0 eV. The Fermi energy sits inside a pseudo-gap region spanning $E_{\text{F}}$ $\pm$ 1 eV with the valence bands being derived from Ge 4$p$, Y 4$d$ and Al 3$p$ orbitals, while the conduction bands mostly from empty Y 4$d$ orbitals. This pseudo-gap region is also reflected in the band dispersion as plotted in Fig. \ref{BandStructure}b without and \ref{BandStructure}d with SOC. Between the highest valence and lowest conduction band, there is a sizable energy separation along most of the high-symmetry directions in the BZ, giving a local DOS minimum and pseudo-gap region near the $E_{\text{F}}$. Without SOC, Fig. \ref{BandStructure}b shows  that crossings along the $\Gamma-X$, $\Gamma-Y$, $Y-X1$, $X1-A1$ and $T-Y$ directions form 4-fold degenerate nodal loops. With SOC in \ref{BandStructure}d, these nodal loops are all gapped out, giving a continuous gap between the highest valence and lowest conduction band, but not a global band gap. This continuous gap between the two band manifolds gives a well-defined topological index. From the Wilson loops plotted in \ref{BandStructure}c, the odd crossings for the Wannier charge centers on both $k_{\text{z}}$ = 0.0 and $k_{\text{z}}$ = 0.5 planes but not on other planes indicate a weak topological insulator phase is hosted in the system, albeit still with a finite Fermi surface, similar to what is observed in RhBi$_2$. \cite{lee2021discovery} The non-magnetic band structures of other RAlGe series are similar. 

To supplement the quantum oscillations observed in the magnetization isotherms, we also used DFT to calculate Fermi surfaces (FS). There are three bands crossing $E_{\text{F}}$, as seen in Fig. \ref{BandStructure}d and their Fermi surfaces are plotted in Figs. \ref{BandStructure}e-h, with the combined FS shown in Fig. \ref{BandStructure}h. The next valence band below the highest energy valence band only crosses $E_{\text{F}}$ near the $A1-T$ direction and forms cone-shaped hole pockets as seen in Fig. \ref{BandStructure}e. The valence band FS in Fig. \ref{BandStructure}f is a skeleton spanning along the $k_y$ (or the $b$-axis) direction. The different sizes of the cross sections of the skeleton give cyclotron extremal orbits for quantum oscillations when $H \parallel b$. Our calculation shows three main frequencies at 30.8, 172.0 and 393.1 T, with the highest frequency agreeing reasonably well with the experimentally measured 354 T pocket in Fig. \ref{YAlGe_QOs}c. The conduction band's FS, shown in Fig. \ref{BandStructure}g, occupies the two ends of the Brillouin zone opposite to the valence band FS, with no overlap between them, confirming a continuous gap. When $H \parallel c$, there are two main cyclotron extremal orbits. One is the cone-shaped FS from the next valence band and the second is from the valence band, which form two concentric extremal orbits near the $A1-T$ direction (see Fig. \ref{BandStructure}h) with the frequencies of 229.0 and 321.1 T, respectively. Given that the average (275 T) of these two frequencies matches very closely the 272 T experimental value with an asymmetric shape as seen in Fig. \ref{BandStructure}d, it is possible this frequency corresponds to scattering between the two 229 T and 321 T calculated pockets. These results and calculations motivate more detailed study, reserved for future work, of the of the quantum oscillations and Hall effect to further study electronic band structure, especially in the magnetically ordered states of moment-bearing members of the \textit{R}AlGe series.

\begin{figure*}[t!]
    \centering
    \includegraphics[width=\linewidth]{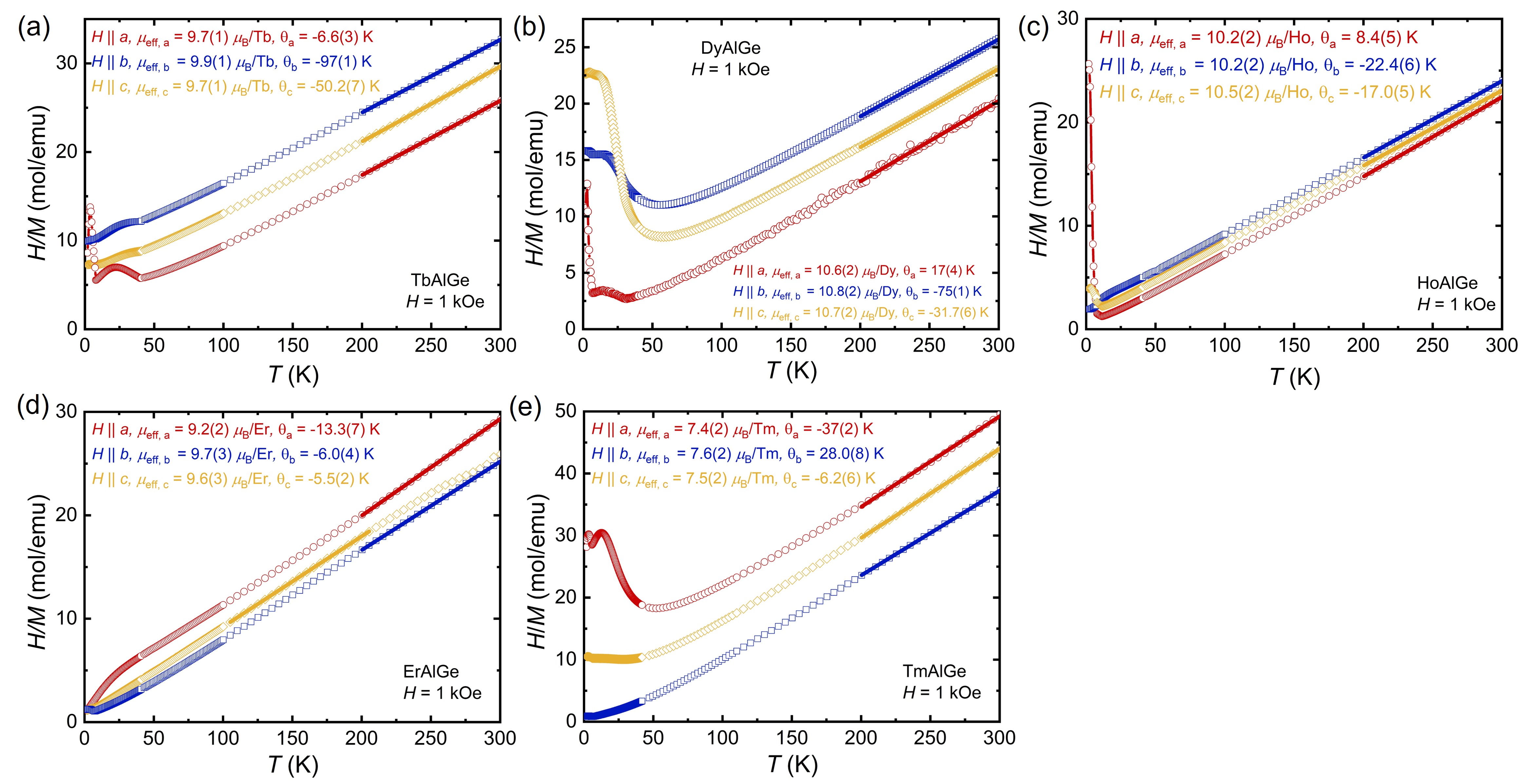}
    \caption[]{Inverse $M/H$ plotted against temperature for $R$AlGe for $R$ = Tb--Tm. The solid lines are Curie-Weiss fits to the respective data. Results for GdAlGe are not shown since Gd has $L$ = 0, leading to negligable CEF effects and isotropic paramagnetic behavior. Note that in ErAlGe, the the high temperature behavior of $(M/H)^{-1}$ deviates from linearity above $\approx$ 225 K for the data measured in the $H \parallel c$ orientation. We believe this is an experimental artifact, which gives clearly erroneous values of $\mu_{eff}$ and $\theta_c$; therefore, we performed Curie-Weiss fits between 100-200 K, where the data is linear with similar slope to the $H \parallel a$ and $H \parallel b$ data.}
    \label{X-1anis}
\end{figure*}

\section{Summary and conclusions}

We present a detailed study of the magnetization, electronic transport, and specific heat of the \textit{R}AlGe family, where \textit{R} = Y, Gd--Lu (except Yb). Whereas the light rare earth \textit{R}AlGe variants (\textit{R} = La--Gd) crystallize in a tetragonal structure, the heavy \textit{R}AlGe materials adopt an orthorhombic \textit{Cmcm} crystal structure. Here, we find that the moment-bearing heavy \textit{R}AlGe materials order antifferomagnetically at \textit{T}$_N$ ranging from 5 K (TmAlGe) to 39 K (TbAlGe). We also find a second, low temperature antiferromagnetic transition occurs at 6--8 K when \textit{R} = Gd--Ho. Magnetization isotherms measured at 2 K reveal multiple metamagnetic transitions (under 70 kOe) for $\textit{H}$ $\parallel$ $\textit{a}$ for \textit{R} = Tb--Ho, whereas a single metamagnetic transition is observed with the field along the \textit{b}-axis when \textit{R} = Er and Tm. While not explored here, the 2 K \textit{M}(\textit{H}) isotherms suggest the \textit{R} = Tb--Ho members will have rich \textit{H}--\textit{T} phase diagrams. Transport measurements show metallic behavior and moderate positive magnetoresistance of 60--150$\%$ at \textit{T} = 2 K. The field dependent magnetization measurements on non-moment bearing YAlGe show clear de Haas–van Alphen oscillations, suggesting small, high mobility pockets at the Fermi level. DFT calculations indicate the nonmagnetic band structure contains nodal loops that are gaped by spin-orbit coupling, supporting a weak topological insulator phase. The overview of physical properties presented here lays the foundation for future, more detailed work on individual \textit{R}AlGe members.

\section{Acknowledgments}

Work at Ames National Laboratory was supported by the U.S. Department of Energy, Office of Science, Basic Energy Sciences, Materials Sciences and Engineering Division. Ames National Laboratory is operated for the U.S. Department of Energy by Iowa State University under Contract No. DE-AC02-07CH11358. Some of the computation used resources of the National Energy Research Scientific Computing Center (NERSC), a DOE Office of Science User Facility. We thank Hahnbidt Rhee for beginning the crystal growth and characterization of the samples discussed in this paper.

\vskip 0.25cm
\noindent
*corresponding authors' email: slade@ameslab.gov, canfield@ameslab.gov

\vskip 0.25cm
\noindent
\textbf{\textit{Data Availability.}} The data that support the findings of this study will be made available in DataShare, an open-access repository at Iowa State University at DOI: 10.25380/iastate.32226210.

\vskip 0.25cm
\noindent
\textbf{\textit{Conflicts of Interest}}

The authors have no conflicts of interest to declare.

\appendix
\section{Estimation of leading crystal field parameters}

As outlined in the discussion section of the main text, the $R$AlGe ($R$ = Tb--Tm) compounds show evidence for strong crystal field effects influencing the magnetic anisotropy and de Gennes scaling of $T_N$. Whereas the full CEF Hamiltonian for orthorhombic ($Cmcm$) $R$AlGe, with rare earth atoms in $m$2$m$ point symmetry, is exceedingly complex, we can use a simplified Hamiltonian to estimate the leading crystal field parameters. If only the second-order terms are assumed to be dominant, an approximate CEF Hamiltonian can be expressed as

\begin{equation}
    H_{CEF} \approx B_2^0O_2^0 + B_2^2O_2^2
\end{equation}

\noindent In this approximation, the CEF parameters can be directly related to the anisotropic Weiss temperatures obtained from Curie-Weiss fits to the magnetic susceptibility as follows \cite{wang1971crystal,shohata1977magnetic,hashimoto1979magnetic}:

\begin{equation}
\begin{split}
    &\theta_a = \theta_{p} + (2J-1)(2J+3)(B_2^0 + B_2^2)/10k_B \\
    &\theta_b = \theta_{p} - (2J-1)(2J+3)B_2^0/5k_B \\
     &\theta_c = \theta_{p} + (2J-1)(2J+3)(B_2^0 - B_2^2)/10k_B
\end{split}
\end{equation}

\noindent where $J$ is the total angular momentum for each respective rare-earth ion and $\theta_{p}$ represents the paramagnetic Weiss temperature from only the molecular field. Here we assume the polycrystalline averaged $M/H$ gives a reasonable estimate of the magnetization without the influence of the crystal field so $\theta_p = \theta_{avg}$. The anisotropic ($M/H$)$^{-1}$ curves and associated Curie-Weiss fits are shown in Fig. \ref{X-1anis}. 

Equations (A.2) provide several routes to estimate the leading crystal field parameters. Using $\theta_b$ to determine $B_2^0$, $B_2^2$ can then be independently estimated using either $\theta_a$ or $\theta_c$. Alternatively, $\theta_a$ and $\theta_c$ can be used to determine $B_2^0$ and $B_2^2$ without use of $\theta_b$. The results of both approaches are plotted in Fig. \ref{dG}c and \ref{dG}d and reveal two main conclusions. First, $B_2^0$ trends monotonically across the series and changes sign from positive for $R$ = Tb-Ho to negative for $R$ = Er and Tm. Second, the absolute value of both parameters, especially $B_2^2$, are poorly determined by this approach, with extreme results from differing calculations differing by an factor of nearly four for TbAlGe and TmAlGe. The wide range of possible values may reflect either unaccounted for uncertainties in the determination of the Weiss temperatures and/or the limitations of using a Hamiltonian truncated at second-order terms.

\begin{figure}[t!]
    \centering
    \includegraphics[width=\linewidth]{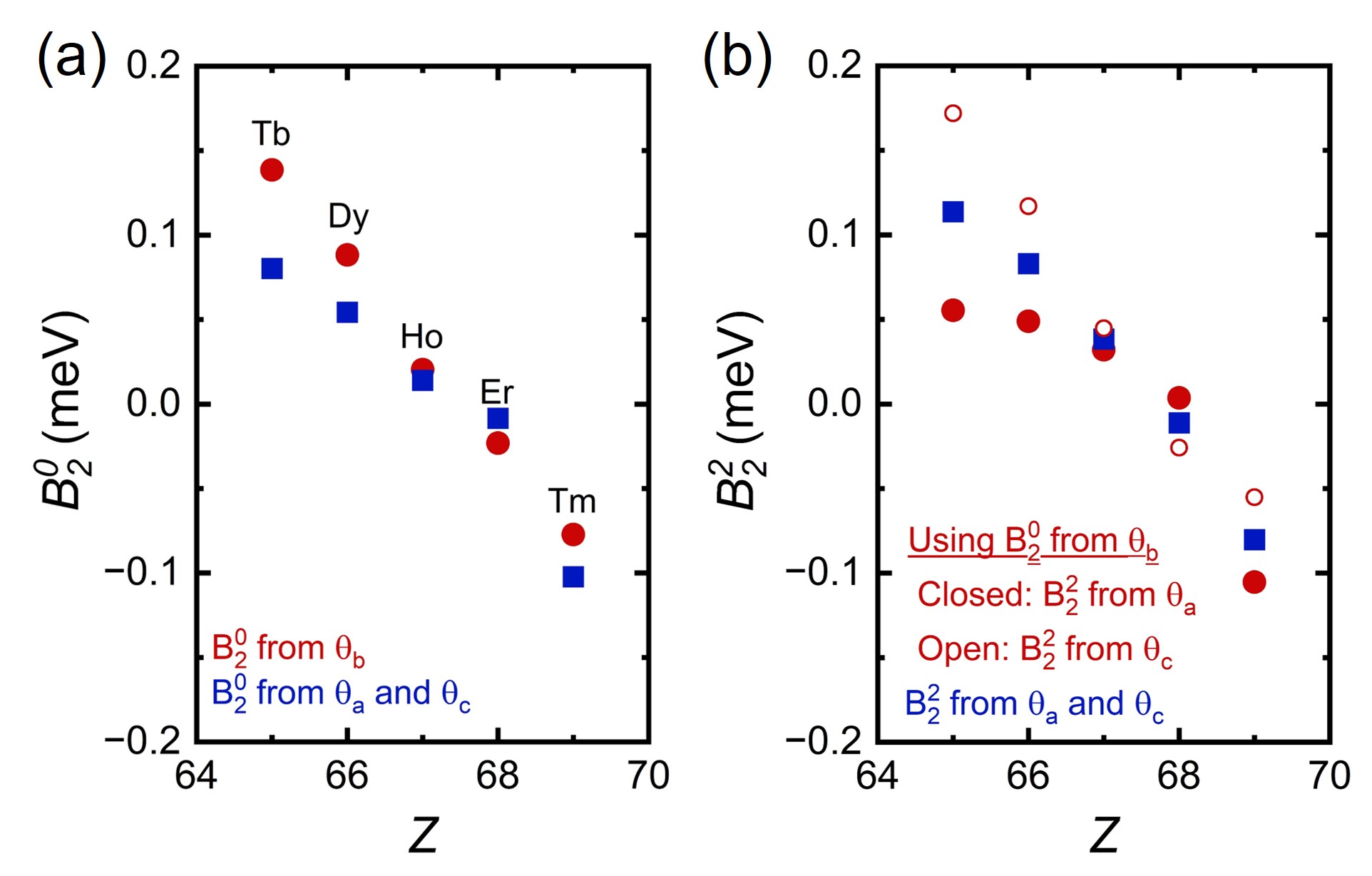}
    \caption[]{Second-order crystal field parameters of the $R$AlGe $R$ = Tb--Tm series estimated using the experimental Weiss temperatures and Eqs. A.2. (a) $B_2^0$ and (b) $B_2^2$.}
    \label{Bnm}
\end{figure}

Despite providing a relatively poor estimate of the absolute values of $B_2^0$ and $B_2^2$, we can still gain useful information from the inferred crystal field parameters. The sign of $B_2^0$ determines whether the system displays an easy $z$-axis (if $B_2^0$ $<$ 0) or easy $xy$-plane (if $B_2^0$ $>$ 0), and $B_2^2$ leads to further asymmetry between the $x-$ and $y$-axes. In the reference frame with the $z$- crystal field axis along the $b$-axis of the $R$AlGe structure, the crystal field's $x$- and $y$-axis are along the respective $c$- and $a$- crystallographic directions. In this basis, a negative $B_2^0$ will therefore favor an easy $b$-axis, whereas if $B_2^0$ is positive, a positive(negative) $B_2^2$ will favor an easy $a$($c$)-axis. Examining Fig. \ref{Bnm} and using any set of the estimated parameters, we find positive $B_2^0$ and $B_2^2$ for $R$ = Tb--Ho samples, indicating an easy $a$-axis, and for $R$ = Er, Tm, a negative $B_2^0$ is found, indicating an easy $b$-axis. These qualitative observations support that the gross features of the magnetic anisotropy in the $R$AlGe ($R$ = Tb--Tm) series are dictated by the crystal field. In particular, the change from having an easy $a$- to $b$- axis on crossing from HoAlGe to ErAlGe is due to the changing sign of the leading $B_2^2$ parameter. 

\section{Crystallographic tables}

\begin{table}[htbp]
  \centering
  \caption{Atomic positions, site occupancies, and isotropic thermal displacement parameters for HoAlGe determined from single crystal X-ray diffraction refinement.}
    \resizebox{\linewidth}{!}{\begin{tabular}{llrrrrl}
    \toprule
    Atom & Site & \multicolumn{1}{l}{x} & \multicolumn{1}{l}{y} & \multicolumn{1}{l}{z} & \multicolumn{1}{l}{Occ.} & U$_{iso}$ (\AA$^2$) \\
    \midrule
    Ho & 4c  & 1   & \multicolumn{1}{l}{0.18991(2)} & 0.75 & 1   & 0.00643(13) \\
    Ge & 4c  & 0.5 & \multicolumn{1}{l}{0.39318(6)} & 0.75 & 1   & 0.00687(15) \\
    Al & 4a  & 0   & 0.5 & 0.5 & 1   & 0.0082(3) \\
    \bottomrule
    \end{tabular}}%
  \label{table_coordinates}%
\end{table}%

\newpage

\begin{table}[htbp]
  \centering
  \caption{Anisotropic thermal displacement parameters (in \AA$^2$) for HoAlGe determined from single crystal X-ray diffraction refinement.}
    \resizebox{\linewidth}{!}
    {\begin{tabular}{lllllrr}
    \toprule
    Atom & $U_{11}$ & $U_{22}$ & $U_{33}$ & $U_{23}$ & \multicolumn{1}{l}{$U_{13}$} & \multicolumn{1}{l}{$U_{12}$} \\
    \midrule
    Ho & 0.00528(16) & 0.00652(17) & 0.00749(18) & \multicolumn{1}{r}{0} & 0   & 0 \\
    Ge & 0.0049(3) & 0.0070(3) & 0.0086(3) & \multicolumn{1}{r}{0} & 0   & 0 \\
    Al & 0.0059(8) & 0.0103(8) & 0.0083(8) & 0.0000(6) & 0   & 0 \\
    \bottomrule
    \end{tabular}}%
  \label{table_Uanis}%
\end{table}%

\begin{table}[htbp]
  \centering
  \caption{Single crystal X-ray diffraction data and structural refinement information for HoAlGe}
    \resizebox{\linewidth}{!}{\begin{tabular}{ll}
    \toprule
    Chemical formula & HoAlGe \\
    Formula weight (g/mol) & 264.5 \\
    Temperature & 297.49(10) \\
    Wavelength (\AA, Ag $K_{\alpha}$) & 0.56087 \\
    Crystal system & orthorhombic \\
    Space group & \textit{Cmcm} (\#63) \\
    \multirow{3}[0]{*}{Unit cell dimensions} & $a$ = 4.0344(3) \AA\, alpha = 90 \\
        & $b$ = 10.3634(7) \AA\, beta = 90 \\
        & $c$ = 5.7290(4) \AA\, gamma = 90 \\
    Volume (cm\^3) & 239.53(3) \\
    $Z$   & 4 \\
    Calculated density (g/ cm$^3$) & 7.335 \\
    Absorption coefficient (mm$^{-1}$) & 24.024 \\
    Absorption Correction & multi-scan \\
    F(000) & 448 \\
    Crystal size (mm) & 0.28 x 0.19 x 0.1 \\
    theta range for data collection & 3.0840-31.537 \\
    Index ranges (min/max, $h$, $k$, $l$) & [-7/7, -18/19, -10/7] \\
    Reflections collected & 2645 \\
    Independent reflections & 436 ($R_{\text{int}}$ = 0.0914) \\
    Completeness & 100 \\
    Refinement method & Full-matrix least-squares on $F^2$ \\
    Data / restrains / parameters & 436 / 0 / 14 \\
    GOF & 1.043 \\
    Final R indices [I $>$ 2$\sigma(I)$] & $R_{obs}$ = 0.0283, $wR_{\text{obs}}$ = 0.0597 \\
    $R_{\text{indices}}$ [all data] & $R_{\text{all}}$ = 0.0293, $wR_\text{all}$ = 0.0606 \\
    Extinction coefficient & 0.078(3) \\
    Largest diff. peak and hole (e-/\AA$^3$) & 3.627 and -3.309 \\
    \bottomrule
    \end{tabular}}
  \label{table_refinement}%
\end{table}%

\bibliographystyle{elsarticle-num} 
\bibliography{references.bib}

\end{document}